\PassOptionsToPackage{dvipsnames,table}{xcolor}
\PassOptionsToPackage{hyphens}{url}
\documentclass[sigplan,10pt,balance=false,nonacm]{acmart}

\setcopyright{none}
\acmDOI{}
\acmISBN{}
\date{}

\makeatletter
\newcommand{\paper@author@display}{%
  \def\@author##1{%
    \ifx\@currentauthors\@empty
      \gdef\@currentauthors{##1}%
    \else
      \g@addto@macro\@currentauthors{,\space##1}%
    \fi
    \gdef\and{}}%
  \let\email\@gobbletwo
  \def\city##1{\global\@ACM@citypresenttrue}%
  \def\country##1{\global\@ACM@countrypresenttrue}%
  \addresses}
\patchcmd{\@mkauthors@iii}
  {\addresses}
  {\paper@author@display}
  {}
  {\PackageError{author-layout}{Cannot adjust ACM author display}{Check the acmart author renderer after a template update.}}
\makeatother

\usepackage{listings}
\usepackage{color}
\usepackage{makecell}    
\usepackage{amsmath,amsopn}                        %
\usepackage{thmtools}

\usepackage{subcaption}                                           %
\usepackage{endnotes,microtype,xspace,fancyvrb,multirow} %
\usepackage{booktabs}                                             %
\usepackage{array,underscore,relsize}                             %
\usepackage{enumitem}                                             %
\newcommand{\paperitemizebefore}{%
  \par\setlength{\parskip}{0pt}\vspace{2pt}}
\setlist[itemize]{topsep=0.25ex,partopsep=0pt,itemsep=1pt,
  before=\paperitemizebefore}
\usepackage[export]{adjustbox}                                    %
\usepackage{pifont}                                               %
\usepackage{tablefootnote}                                        %
\usepackage{graphicx}
\usepackage[linesnumbered,vlined,ruled,commentsnumbered]{algorithm2e}
\DontPrintSemicolon
\newcommand{\ra}[1]{\renewcommand{\arraystretch}{#1}}
\SetCommentSty{mycommfont}

\usepackage{hhline}
\usepackage{pgf}
\usepackage{soul}
\usepackage{contour}

\definecolor{tablegray}{gray}{0.85}

\usepackage[normalem]{ulem}

\usepackage{textcomp}
\usepackage{graphicx}

\usepackage[compact]{titlesec}
\titleformat*{\section}{\large\bfseries}
\titleformat*{\subsection}{\normalsize\bfseries}
\titleformat{\subsubsection}[hang]{\normalsize\bfseries}{\thesubsubsection}{1em}{}
\titlespacing*{\section}{0pt}{*1.3}{*0.8}
\titlespacing*{\subsection}{0pt}{*1.}{*0.5}
\titlespacing*{\subsubsection}{0pt}{1.3ex}{0.8ex}

\makeatletter
\AtEndPreamble{%
  \setlength{\parskip}{0pt plus 1pt}%
  \setlength{\partopsep}{2pt plus 1pt minus 1pt}%
  \setlength{\floatsep}{11pt plus 2pt minus 2pt}%
  \setlength{\textfloatsep}{18pt plus 2pt minus 4pt}%
  \setlength{\intextsep}{11pt plus 2pt minus 2pt}%
  \setlength{\dblfloatsep}{11pt plus 2pt minus 2pt}%
  \setlength{\dbltextfloatsep}{18pt plus 2pt minus 4pt}%
  \setlength{\skip\footins}{8.1pt plus 4pt minus 2pt}%
  \setlength{\@fptop}{0pt plus 1fil}%
  \setlength{\@fpsep}{8pt plus 2fil}%
  \setlength{\@fpbot}{0pt plus 1fil}%
  \setlength{\@dblfptop}{0pt plus 1fil}%
  \setlength{\@dblfpsep}{8pt plus 2fil}%
  \setlength{\@dblfpbot}{0pt plus 1fil}%
}
\makeatother

\AtEndPreamble{%
  \captionsetup{font=small,labelfont=bf,textfont=normalfont,
    labelsep=period,skip=1pt}%
  \setlength{\belowcaptionskip}{-6pt}%
}

\definecolor{mygreen}{rgb}{0,0.6,0}  
\definecolor{mygray}{rgb}{0.5,0.5,0.5}  
\definecolor{mymauve}{rgb}{0.58,0,0.82}

\definecolor{paperurlcolor}{HTML}{005595}
\definecolor{papercitecolor}{HTML}{781D7D}
\AtEndPreamble{%
  \hypersetup{
    colorlinks=true,
    linkcolor={red!50!black},
    citecolor=papercitecolor,
    urlcolor=paperurlcolor
  }
}

\newcommand{\sys}{\emph{\texttt{libumsh}}\xspace}
\newcommand{\Bind}{{\textsc{Bind}}\xspace}
\newcommand{\Write}{{\textsc{Write}}\xspace}
\newcommand{\Read}{{\textsc{Read}}\xspace}
\newcommand{\Unbind}{{\textsc{Unbind}}\xspace}
\newcommand{\Acquire}{{\textsc{Acquire}}\xspace}
\newcommand{\Release}{{\textsc{Release}}\xspace}

\newcommand{\nostitle}[1]{\noindent{\bf #1}}
\newcommand{\stitle}[1]{\vspace{0.8ex}\noindent{\bf #1}}
\newcommand{\etitle}[1]{\vspace{0.5ex}\noindent{\em\ul{#1}}}

\newcommand{\inst}[1]{\texttt{#1}}

\newcommand{\squishlist}{
  \begin{list}{$\bullet$}{
    \setlength{\itemsep}{0pt}
    \setlength{\parsep}{3pt}
    \setlength{\topsep}{3pt}
    \setlength{\partopsep}{0pt}
    \setlength{\leftmargin}{1.5em}
    \setlength{\labelwidth}{1em}
    \setlength{\labelsep}{0.5em}
  }
}

\newcommand{\squishend}{
  \end{list}
}

\usepackage{tikz}

\usepackage{upgreek}

\title{Characterizing CPU-GPU Memory Sharing over Unified Memory Architecture}

\author{Jinrong Yang}
\email{coyangjr@sjtu.edu.cn}
\author{Zimeng Wang}
\email{simonwzm@sjtu.edu.cn}
\author{Rong Chen}
\authornote{Rong Chen is the corresponding author (\url{rongchen@sjtu.edu.cn}).}
\email{rongchen@sjtu.edu.cn}
\author{Haibo Chen}
\email{haibochen@sjtu.edu.cn}
\affiliation[obeypunctuation=true]{%
  \department{Institute of Parallel and Distributed Systems},
  \department{School of Computer Science},
  \institution{Shanghai Jiao Tong University}%
  \city{Shanghai}\country{China}}
\renewcommand{\shortauthors}{Jinrong Yang et al.}

\makeatletter
\let\ACM@origsection\section
\let\ACM@origsubsection\subsection
\let\ACM@origsubsubsection\subsubsection
\let\ACM@origparagraph\paragraph
\let\ACM@origbaselinestretch\baselinestretch
\makeatother

\acmBadgeR[https://www.acm.org/publications/policies/artifact-review-and-badging-current]{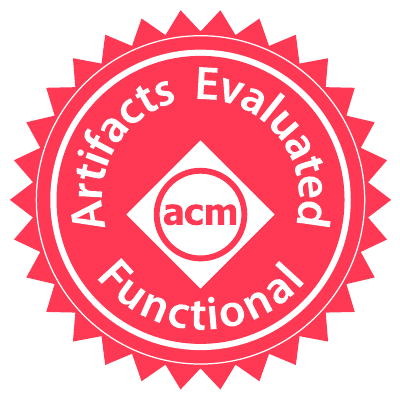}
\acmBadgeR[https://www.acm.org/publications/policies/artifact-review-and-badging-current]{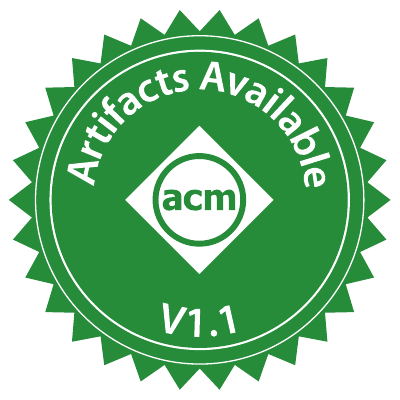}
\makeatletter
\let\paper@badges\@acmBadgeR
\def\@acmBadgeR{\raisebox{-40pt}{\paper@badges}\hspace*{-20pt}}
\makeatother

\begin{document}

\frenchspacing

\begin{abstract}

Unified Memory Architecture (UMA) enables the CPU and GPU to share a single physical memory. 
However, its complexity and opacity often drive applications back to inefficient copying 
or to incorrect direct access that sacrifices performance or even correctness. 
This paper presents a comprehensive study of CPU-GPU memory sharing on UMA, 
covering correctness and performance across data-plane transmission and 
control-plane coordination on four platforms.
It yields four findings that guide {\sys}, a portable library that delivers
correct and efficient memory sharing across platforms and extends to 
emerging ones (e.g., NVIDIA DGX Spark).
Applied to nine real-world systems, 
{\sys} reduces data-plane time by 16–75\% and control-plane overhead by 7–90\%.

\end{abstract}

\maketitle

\section{Introduction}
\label{sec:intro}

\begingroup
\brokenpenalty=100\relax
\doublehyphendemerits=50000\relax
Heterogeneous SoCs that couple a CPU with one or more 
accelerators---GPUs, NPUs, or ASICs---have become the dominant substrate 
for model inference, graphics rendering, scientific computing, and generative AI. 
Across edge and client devices, including embedded systems~\cite{raspberrypiRaspberryPi52025,
nvidiacorporationNVIDIAJetsonAGX2025,huaweiHuaweiMDC6102025,
appleincAppleWatch2025}, 
mobile phones~\cite{qualcommSnapdragonSmartphones2025,
appleincIPhone2025a,mediatekincMediaTekDimensitySmartphones2025}, 
laptops~\cite{intelcorporationIntelCoreUltra2025,appleincMacBookPro2025,
qualcommLaptopsPoweredSnapdragon}, 
and emerging mini AI supercomputers~\cite{skende171NVIDIAGB102026,
beelinkBeelinkGTR9Pro2025,appleincMacStudio2025}, vendors increasingly integrate 
these accelerators with the CPU under a Unified Memory Architecture (UMA),
in which both processors share a single physical memory. 
In contrast to the traditional Discrete Memory Architecture, which
keeps CPU and GPU memory physically separate and connects them 
over PCIe or NVLink C2C~\cite{nvidiacorporationNVIDIAGraceHopper2025}, 
UMA offers a simpler programming model, a larger memory footprint available 
to each processor, and the elimination of explicit cross-device copies.
\par\endgroup

This paper studies CPU-GPU communication via memory sharing 
on UMA platforms---a basic building block that determines both the correctness
and the performance of heterogeneous computing.
Such communication serves two qualitatively distinct purposes. 
Data-plane transmission moves bulk data into or out of 
GPU computation, including model parameters, inference inputs/outputs, textures, 
and point clouds, typically at task boundaries;
representative scenarios include model loading~\cite{pytorchPyTorch,
intelcorporationOpenVINODocumentation,ggmlorgLlamacpp2025}, 
image preprocessing~\cite{opencvteamOpenComputerVision2025,ultralyticsincUltralytics2025}, 
output sampling~\cite{openvinotoolkitOpenVINOGenAI2025,transformers}, and 
frame composition~\cite{khronosgroupMoltenVK2025,khronosgroupincVulkan2025}. %
Control-plane coordination exchanges small but frequent 
state---flags, counters, and locks---within and across kernels to sequence work 
or arbitrate shared resources between the CPU and GPU,
as exemplified by hybrid inference~\cite{xue2024powerinfer,
gomezlunaChaiCollaborativeHeterogeneous2017a,zhangEdgeNNEfficientNeural2023a,
chenCharacterizingMobileSoC2025}, 
model offloading~\cite{HuggingfaceAccelerate2025,
leeInfiniGenEfficientGenerative2024,chenKTransformersUnleashingFull2025}, 
and GPU filesystem operations~\cite{silberstein2014gpufs,
qiuGeminiFSCompanionFile2025,liManagingScalableDirect2025}. 

Although UMA platforms natively support \emph{direct access} to shared memory,
mainstream applications routinely fail to exploit this capability correctly or efficiently.
In practice, most applications 
either treat the platform as 
if it were discrete and incur unnecessary copy overhead~\cite{ggmlorgLlamacpp2025,ollamaOllama,
vllmprojectvllm2025,deepspeedaiDeepSpeed2025,pytorchPyTorch,tensorflowTensorFlow2015,
nvidiaTensorRT2025,intelcorporationOpenVINODocumentation}, 
or apply UMA features incorrectly and suffer data inconsistency or 
crashes~\cite{khronosgroupmoltenvkSupportPermanentlyOpen2018,
khronosgroupmoltenvkLinearImagesCant2019,nvidiadeveloperforumsDualProblemsUnified2020,
nvidiadeveloperforumsUnifiedMemoryConcurrency2025}.
For example, widely used deep learning frameworks 
such as PyTorch~\cite{pytorchPyTorch} and OpenVINO~\cite{intelcorporationOpenVINODocumentation},
as well as LLM serving systems 
such as vLLM~\cite{vllmprojectvllm2025} and llama.cpp~\cite{ggmlorgLlamacpp2025},
still rely on redundant copies for weight loading and I/O processing 
on UMA platforms, wasting both time and memory.
Rendering engines such as MoltenVK~\cite{khronosgroupMoltenVK2025} %
exhibit incorrect usage patterns that cause data inconsistency between the CPU and GPU.

These failures are not matters of engineering preference; they arise from 
the inherent complexity and opacity of UMA platforms along three aspects. 
(i) \emph{Hardware}: platforms differ in interconnect type (Ring vs. NoC), 
in the presence of memory-side cache (MSC) in front of DRAM, and 
in support for GPU-to-CPU coherence, and each combination alters which 
access patterns are correct and which are efficient. 
(ii) \emph{Software}: critical configurations such as page size, cache policy, 
and coherence direction are buried within closed-source drivers and 
proprietary runtimes, while the OS abstractions above them silently fall back 
under load~\cite{kwonIngensHugePage2017,panwarMakingHugePages2018}. 
(iii) \emph{Semantics}: data-plane and control-plane traffic impose opposing 
objectives---bandwidth on large transfers versus low-contention atomics 
on small flags---so no single memory-sharing recipe is universally optimal.

\begin{figure}[t]
  \vspace{2mm}
  \begin{minipage}{1.\linewidth}
    \centering\includegraphics[width=\linewidth,viewport=62 509 533 692,clip=true]{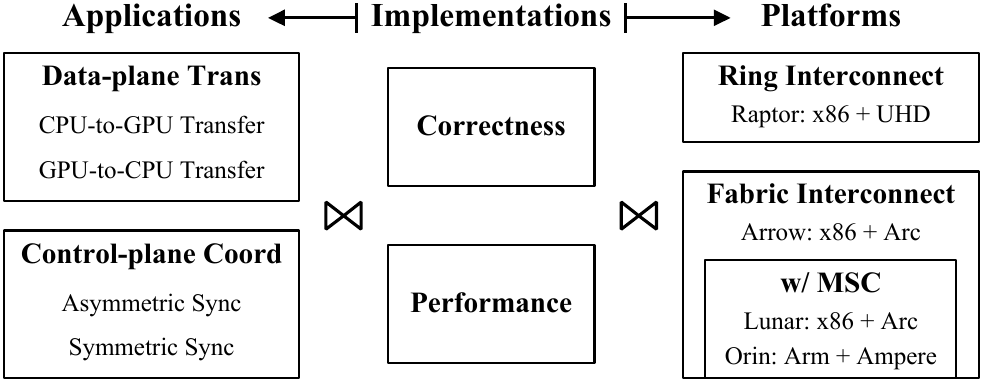}
  \end{minipage} \\[4pt]
  \begin{minipage}{1.\linewidth}
    \caption{\small{\textit{Our analysis methodology.}}}
    \label{fig:method}
  \end{minipage}  \\[-8pt]
\end{figure}

In response, 
we present the most comprehensive study of CPU-GPU memory sharing over UMA to date. 
We pursue completeness along three dimensions (Figure~\ref{fig:method})
that mirror the source of complexity above:
applications (semantics), implementations (software), and platforms (hardware).

\begin{itemize}[leftmargin=2.3ex,topsep=0.25ex]
\item \textbf{Applications:}
We classify the purposes of CPU-GPU memory sharing into two planes: 
\emph{data-plane transmission}, which moves data for GPU computation and covers both
\emph{CPU-to-GPU} and \emph{GPU-to-CPU} transfer; 
and \emph{control-plane coordination}, which exchanges state information between 
the CPU and GPU and covers both \emph{asymmetric} and \emph{symmetric} synchronization.

\item \textbf{Implementations:}
We decompose every sharing operation into an independently configurable 
\emph{sending} and \emph{receiving} stage, each tunable along 
three attributes (page size, cache policy, coherence) and realizable as 
Write-Alone, Write-then-Flush, Write-Bypass, or their read counterparts.
This allows us to enumerate every correct combination per platform and 
identify the fastest for the data and control planes.

\item \textbf{Platforms:}
We evaluate four x86/ARM platforms with Intel/NVIDIA GPUs,
representing UMA designs used from cost-sensitive devices to high-performance
systems. They contrast Ring and Fabric NoC interconnects,
memory-side caches (MSC), and one-way cache coherence (CC).
Spark extends the study to two-way CC (\S\ref{sec:impl-spark}).

\end{itemize}

\stitle{Findings.} 
Our analysis covers what we believe are the most representative CPU-GPU 
memory sharing applications and UMA platforms in use today, 
yielding correct implementations for each platform and four performance findings
(\S\ref{sec:sharing}). 

\vspace{0.8mm}
\noindent {1.}
\etitle{Page size dominates setup cost and propagates into access cost}.~
For large regions, 4\,KB pages inflate setup time by nearly 100$\times$ 
over 2\,MB pages and degrade subsequent bandwidth through TLB pressure. 
For small regions, forcing 2\,MB allocations wastes memory and is up to 5$\times$ 
slower on Lunar and Orin. 
Transparent Huge Pages (THP)~\cite{linuxkerneldevelopmentcommunityTransparentHugepageSupport2025} 
selects an appropriate size automatically; 
we therefore recommend enabling THP and prefaulting the region at setup.

\vspace{0.8mm}
\noindent {2.}
\etitle{Uncached memory is simple and platform-agnostic, 
but its cost is substantial}.~ 
Uncached (UC) memory makes direct access correct without code changes---as with
Orin's default GPU policy for registered host-memory
sharing~\cite{nvidiacorporation1CUDATegra2025}---yet
it inflates single-threaded CPU latency by up to 300$\times$. 
GPU parallelism can hide UC overhead in microbenchmarks, but real workloads still 
incur a measurable penalty: an L2-optimized GEMM on Orin loses 16\% of its GFLOPS. 
UC should therefore be reserved for cases in which
neither code modification nor coherence support is feasible.

\vspace{0.8mm}
\noindent {3.}
\etitle{One-way cache coherence simplifies implementation 
at near-zero cost on MSC-enabled platforms}.~ 
GPU-to-CPU cache coherence (CC), supported on most UMA platforms,
allows the CPU to correctly access write-back (WB) memory 
without any application-level changes.
CC introduces overhead on GPU accesses, 
but this cost is negligible when a memory-side cache (MSC) is present.
On Arrow, which lacks an MSC, GPU write time with CC is
2.07$\times$ that without CC.

\vspace{0.8mm}
\noindent {4.}
\etitle{Under high concurrency, control-plane coordination is bottlenecked 
by contention rather than access latency}.~
For small shared regions, the choice among correct implementations is
nearly irrelevant; what matters is the number of GPU threads that touch 
the flag concurrently. 
Block-level barriers handle up to 1,000 threads with minimal code complexity. 
Beyond this threshold, grid-level barriers eliminate contention at the cost of 
deadlock-avoidance complexity, and a single-thread kernel remains a safe fallback.

\stitle{Library and extensibility.} 
{\sys} automatically selects correct, efficient implementations through
high-level sender-receiver operations, based on the platform and application
sharing requirements
(Table~\ref{tab:suggested-impl}, \S\ref{sec:impl}).
An AI-assisted three-step porting workflow (\S\ref{sec:impl-extension}) extends
it to NVIDIA DGX Spark~\cite{skende171NVIDIAGB102026}, validating extensibility
to two-way coherence.

\stitle{Case study.} 
We applied {\sys} to
nine real-world systems that span data-plane and control-plane patterns 
across the four target UMA platforms. On the data plane, {\sys} eliminates
redundant copies in model loading (PyTorch), image preprocessing (OpenCV), 
output sampling (OpenVINO GenAI), and point-cloud processing, 
reducing transmission time by 16--75\%.
On the control plane, {\sys} replaces heavyweight runtime events and CPU-side delegation
with lightweight memory-based signaling and atomics in 
hybrid inference (EdgeNN), model offloading (Accelerate, Demand Layering), 
GPU filesystem operations (GPUfs), and cross-device file locking (GoFS),
reducing coordination overhead by 7--90\%.

\stitle{Contributions.}
This paper makes three contributions:
\begin{itemize}[leftmargin=2.3ex,topsep=0.25ex]
    \item We develop a systematic methodology spanning application
    semantics, memory-sharing implementations, and UMA properties
    (Figure~\ref{fig:method}).
    \item We characterize four UMA platforms and reveal
    how page size, cache policy, coherence, MSC, and contention
    jointly shape data-plane and control-plane performance.
    \item We build {\sys}, to our knowledge the first library whose high-level
    CPU--GPU sharing APIs automatically select correct, efficient implementations
    by platform and application sharing requirements.
    The library, correctness tests, and benchmarks are available at
    \url{https://github.com/xpuos/libumsh}.
\end{itemize}

\section{Background and Context}
\label{sec:bg}

\begin{figure*}[t]
  \vspace{2mm}
  \begin{minipage}{1.\linewidth}
    \centering\includegraphics[width=\linewidth,viewport=27 291 564 401,clip=true]{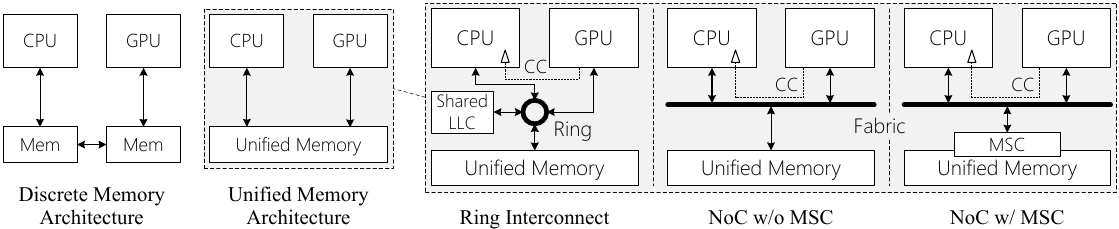}
  \end{minipage} \\[5pt]
  \begin{minipage}{1.\linewidth}
    \caption{\small{\emph{Discrete memory architecture, unified memory architecture (UMA),
	and UMA with different hardware features.
    All interconnect panels are logical views; 
    e.g., the shared LLC on the ring is physically sliced 
    across ring stops co-located with cores.}}}
    \label{fig:march}
  \end{minipage}
\end{figure*}

As heterogeneous computing continues to evolve and the demand for 
high-performance applications increases, various accelerators 
such as GPUs and NPUs are increasingly being deployed 
alongside CPUs. 
This paper focuses on the most representative CPU-GPU communication mechanisms. 

\subsection{Unified Memory Architecture}

Unified memory architecture (UMA) 
enables the CPU and GPU to share a unified physical memory~\cite{seoFACILFlexibleDRAM2025}.
UMA is widely deployed in edge devices, including 
embedded systems~\cite{raspberrypiRaspberryPi52025,nvidiacorporationNVIDIAJetsonAGX2025,huaweiHuaweiMDC6102025,appleincAppleWatch2025}, 
mobile phones~\cite{qualcommSnapdragonSmartphones2025,appleincIPhone2025a,mediatekincMediaTekDimensitySmartphones2025}, 
laptops~\cite{intelcorporationIntelCoreUltra2025,appleincMacBookPro2025,qualcommLaptopsPoweredSnapdragon}, 
and emerging mini AI supercomputers~\cite{skende171NVIDIAGB102026,beelinkBeelinkGTR9Pro2025,appleincMacStudio2025}, 
due to its simplified programming abstraction and increased available memory. 
In contrast, discrete memory architecture,
the traditional mainstream approach, uses 
physically separate CPU and GPU memory spaces connected by
buses such as PCIe or NVLink C2C~\cite{nvidiacorporationNVIDIAGraceHopper2025}.
Figure~\ref{fig:march} illustrates these two memory design approaches. 
With discrete memory, CPU-GPU communication requires explicit data copying
(I/O operations)~\cite{beriUnicornRuntimeEfficient2017, limZicoEfficientGPU2021};
in UMA, all processors can directly access the same data without additional
data movement. 
However, this approach requires a deep understanding of UMA hardware features
and careful coordination between CPU and GPU 
to ensure both correctness and optimal performance.

\stitle{UMA hardware features.}
UMA is deeply integrated into SoC internals, so interconnect designs vary
across vendors and even chip models (Figure~\ref{fig:march}).

\etitle{CPU-GPU Interconnection.}
The studied UMA platforms use a Ring or network-on-chip (NoC) fabric interconnect.
For Ring interconnects, CPU cores and the GPU share a distributed last-level cache (LLC)
accessed over the ring~\cite{paccagnellaLordRingsSide2021, intelRaptorSmartCache2024}.
This requires minimal SoC design changes but scales poorly.
NoC connects them at the cluster level,
scaling better to additional components~\cite{akgunScalableMemoryFabric2016}.
In our NoC platforms, CPU caches are not directly accessible to the GPU
as a shared cache tier.
Some NoC platforms introduce a shared system-level cache above main memory
(memory side cache, MSC)~\cite{croninExplorationARMSystemLevel2021, bonenLunarLakeIntel2025a}.
These cache-sharing and coherence properties depend on platform design,
rather than interconnect topology alone.

\etitle{Data Coherence.}
Cache coherence (CC) between the CPU and GPU provides a consistent shared-memory
view and simplifies UMA programming.
On Ring-interconnect platforms, the GPU, CPU cores, and the LLC all function 
as agents on a shared ring bus~\cite{intelcorporationIntelProcessorGraphics2023}. 
This enables them to participate in a unified cache coherence protocol
that naturally extends the CPU's own inter-core coherence mechanism, 
such as MESIF, although GPU private caches generally 
remain non-coherent with the CPU~\cite{geraPerformanceCharacterisationSimulation2018}.
This coherence is managed automatically and 
is not exposed to the driver for configuration~\cite{linuxI915GemObjectTypeshDriversGpu}.

The studied NoC platforms use
I/O coherency~\cite{lumppFrameworkOptimizingCPUiGPU2021,
bonenLunarLakeIntel2025a}, allowing the GPU to issue coherency transactions 
to the CPU subsystem. 
Coherent GPU reads can retrieve the latest data from CPU caches;
coherent GPU writes invalidate stale CPU cache copies as
required~\cite{nvidiacorporation1CUDATegra2025, AMBACHIArchitecture2025}.
All four target platforms provide only \textit{one-way coherence}: GPU-initiated
accesses can maintain consistency with CPU caches, 
but the CPU cannot reciprocally snoop or invalidate GPU caches.
The cost of GPU-to-CPU coherence depends on the coherence implementation
and cache organization. Raptor uses the ring's shared-LLC coherence protocol,
while the studied NoC platforms use I/O-coherency transactions with the CPU
subsystem that can add memory-access overhead (\S\ref{sec:performance}).

\subsection{CPU-GPU Communication}

CPU-GPU communication is a key building block for heterogeneous computing.
It spans two purposes:
\emph{data-plane transmission} and \emph{control-plane coordination}.\footnote{Our
communication microbenchmarks measure CPU--GPU memory communication without modeling
application-level caching or reuse, such as hot-data locality within a
matrix-multiplication kernel.}

\stitle{Data-plane transmission.}
This covers producer-consumer transfers between CPU and GPU,
including model parameters, inputs, and outputs.
For example, during model inference, data is transferred to and from the GPU 
at the start and end of task execution. Similar transfers occur in 
GPU rendering, 3D image processing, scientific computing, and autonomous 
driving.
Data-plane transmission has two types: 
\textit{CPU-to-GPU} and \textit{GPU-to-CPU} transfer. 
The volume of transferred data varies considerably---ranging from 
a few integers for simple inputs to gigabyte-scale model weights. 
The communication pattern is characterized by low frequency, 
as data transfers generally occur only at GPU task boundaries. 

\stitle{Control-plane coordination.} 
This involves the exchange of control information between the CPU and GPU,
including synchronization signals and status updates in irregular or
CPU--GPU collaborative workloads. For example,
in model offloading scenarios, the CPU iteratively prepares data 
for the GPU---such as loading per-layer parameters or 
a KV Cache---before launching the corresponding kernel.
This unidirectional synchronization pattern is 
referred to as \textit{asymmetric synchronization}.
It can be effectively managed through a wait-based mechanism for 
task dependencies.
Furthermore, 
some heterogeneous workloads (e.g., GPU filesystem) require 
more complex coordination like cross-device contention control, 
where both the CPU and GPU must wait for each other
at specific execution points to ensure correct operation sequencing.
This requires 
advanced synchronization primitives (barriers or locks) 
to enable CPU-GPU coordination. 
This pattern is referred to as \textit{symmetric synchronization}.
To maintain tight coordination between the CPU and GPU during task execution, 
control-plane coordination involves small payloads (typically just 
control messages or flags) yet occurs much more frequently within GPU tasks.

\subsection{Memory Sharing}

CPU-GPU communication can be modeled as one processor (the \emph{sender}) 
transmitting a unit of data to another processor (the \emph{receiver}),
where either the CPU or GPU can act as the sender or receiver. 
Memory sharing between them enables 
CPU-GPU communication. 
Two primary approaches exist for memory sharing: 
\emph{data copying} and \emph{direct access}.

\begin{figure}[t]
  \vspace{1mm}
  \begin{minipage}{1.\linewidth}
    \centering\includegraphics[width=\linewidth,viewport=84 179 327 257,clip=true]{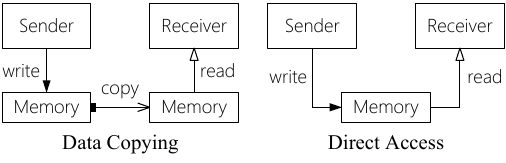}
  \end{minipage} \\[5pt]
  \begin{minipage}{1.\linewidth}
    \caption{\small{\emph{Two approaches for CPU-GPU communication.
    }}}
    \label{fig:comm}
  \end{minipage}  \\[-5pt]
\end{figure}

\stitle{Data copying.}
Analogous to \emph{pass-by-value} in programming, this approach implements 
memory sharing through explicit copy operations between two individual memory regions. 
As shown in Figure~\ref{fig:comm}, the sender first writes data to 
a designated memory region, and then the receiver copies this data into
its own memory region and reads it. 
This approach imposes minimal demands on underlying hardware capabilities, 
making it broadly applicable to both discrete and unified memory architectures.
However, this approach incurs overhead from additional memory copy operations 
and increases memory footprint by maintaining duplicate data copies.

\stitle{Direct access.}
This approach, akin to \emph{pass-by-reference}, 
utilizes a physically shared memory region accessible to 
both sender and receiver, as shown in Figure~\ref{fig:comm}. 
It eliminates copy operations and extra memory consumption.
However, it requires direct CPU and GPU access to the same physical memory
region, a capability provided by UMA platforms.

\stitle{In the wild.}
Despite native support for direct access on UMA platforms, 
many existing systems fail to fully leverage this capability.
Our analysis of popular open-source LLM inference frameworks~\cite{park2025survey},
including llama.cpp~\cite{ggmlorgLlamacpp2025}, Ollama~\cite{ollamaOllama}, 
vLLM~\cite{vllmprojectvllm2025}, and DeepSpeed~\cite{deepspeedaiDeepSpeed2025},
as well as general-purpose model inference frameworks,
including PyTorch~\cite{pytorchPyTorch}, TensorFlow~\cite{tensorflowTensorFlow2015}, 
TensorRT~\cite{nvidiaTensorRT2025}, and OpenVINO~\cite{intelcorporationOpenVINODocumentation},
reveals that these frameworks predominantly rely on data copying
for weight loading and input/output processing.

GPU runtimes provide APIs for direct access, and
developers of rendering and computation frameworks 
use them to optimize performance on UMA platforms.
However, direct access is easy to implement incorrectly: 
missing cache flushes and memory barriers
often result in data inconsistency between CPU and GPU, or application 
crashes~\cite{khronosgroupmoltenvkSupportPermanentlyOpen2018,
khronosgroupmoltenvkLinearImagesCant2019,nvidiadeveloperforumsDualProblemsUnified2020,
nvidiadeveloperforumsUnifiedMemoryConcurrency2025}.
Even with correct implementation, 
suboptimal approaches can greatly impact performance. 
For example, inadequate understanding of memory access characteristics 
may cause direct access to underperform 
data copying~\cite{nvidiadeveloperforumsZeroCopyManagedMemory2016}. 
The model loading optimization for Apple UMA platforms in llama.cpp 
does not effectively bypass the copy operation.
Similarly, improper synchronization granularity often introduces additional overhead 
in control-plane coordination~\cite{osdi2022reef,liManagingScalableDirect2025}.

\stitle{Root causes and challenges.} 
These problems stem from two factors. 
First, these systems were initially designed for discrete memory architectures. 
To reduce development effort, existing data copying implementations 
are directly adapted for UMA platforms. 
Second, substantial variations in hardware architecture, 
software interfaces, and performance across 
UMA platforms~\cite{shen2025xsched} mean that developers often lack 
the experience needed to efficiently implement memory sharing 
via direct access~\cite{lumppFrameworkOptimizingCPUiGPU2021,
henryHighPerformanceDeepLearningCoprocessor2020,bakitaEnablingGPUMemory2022a}. 
Achieving correctness and optimal performance on UMA platforms 
requires precise control over data visibility and 
consistency---a significant challenge in CPU-GPU heterogeneous environments.

\section{Target Platforms}
\label{sec:target}

Our core study covers four platforms with one-way CC.
The x86 Intel subset---Raptor, Arrow, and Lunar---spans Ring, NoC without MSC,
and NoC with MSC, enabling comparisons within one driver/runtime ecosystem.
Orin adds an ARM-based NVIDIA platform with NoC and MSC.

\subsection{Ring Interconnect: Raptor}

The Raptor platform is built around an Intel Core i5-14500 (Raptor Lake) 
processor with
a Ring interconnect 
linking the CPU cores, GPU, and memory controller.
Its 24\,MB LLC is distributed across ring-connected slices and shared by
CPU cores and the GPU~\cite{intelRaptorSmartCache2024}.
The system has dual-channel DDR4-3200 memory 
and an integrated Intel UHD 770 GPU.
Both the GPU and memory controller are integrated as agents on the ring bus.
Memory requests leaving the GPU traverse the ring nodes hosting the LLC's
coherence agents, enabling coherent access to the shared LLC and memory.
However, this coherence is one-way, as the GPU's private 
caches remain non-coherent with the CPU~\cite{geraPerformanceCharacterisationSimulation2018}.
The ring organization remains relevant in low-power, cost-sensitive
products such as Intel Processor N-series
(e.g., N150 and N250)~\cite{intelNSeriesDatasheet2025, intelNSeriesProductBrief2025}.

\subsection{NoC Interconnect without MSC: Arrow}

The Arrow platform uses an Intel Core Ultra 285H (Arrow Lake) processor
with a scalable fabric NoC interconnect. 
The CPU, integrated Intel Arc 140T GPU, and other agents connect
at the cluster level. This architecture offers better scalability.
The CPU's LLC is shared within the compute tile and is not directly
accessible to the GPU as a shared cache tier~\cite{intelArrowSmartCache2025}.
The system includes dual-channel DDR5-5600 memory.
Coherence transactions across the fabric maintain data consistency.
This is also one-way: GPU-initiated accesses can maintain
consistency with CPU caches, but not vice versa.
The absence of an MSC makes this platform essential for
evaluating how a system-level shared cache impacts performance.
The NoC-without-MSC organization also appears in newer embedded devices,
such as AMD Versal Prime Series Gen 2 models 2VM3104 and
2VM3654~\cite{amdVersalOverview2026}.

\subsection{NoC Interconnect with MSC: Lunar and Orin}

Lunar and Orin platforms feature a NoC interconnect
and include an MSC, with one-way coherence.
Apple SoCs~\cite{croninExplorationARMSystemLevel2021, xuEXAMExploitingExclusive2025}
and AMD APUs~\cite{wahlgren2025dissecting} also adopt this organization,
but some coherence differences will be discussed in \S\ref{sec:discussion}.

\stitle{Lunar.}
The Lunar platform, powered by an Intel Core Ultra 258V (Lunar Lake) processor, 
targets energy-efficient mobile computing with graphics and AI acceleration.
The system includes an Intel Arc 140V GPU and LPDDR5-8533 memory. 
Its key feature is an 8\,MB MSC---a cache logically located
between the processors and main memory~\cite{bonenLunarLakeIntel2025a}.
The MSC reduces main memory access delays.

\stitle{Orin.}
The Orin platform is a Jetson AGX Orin embedded system,
built on an ARM Cortex A78AE CPU with an integrated NVIDIA Ampere GPU, 
and demonstrates these architectural trends in the ARM ecosystem. 
It also uses a scalable NoC interconnect and includes a 4\,MB MSC 
to improve memory performance. By including it, 
we conduct a broader analysis of UMA behaviors beyond x86 architecture, 
providing a more comprehensive understanding of the underlying principles.

\section{Communication via Memory Sharing}
\label{sec:sharing}

Analyzing the correctness and performance of CPU-GPU communication 
over UMA presents significant challenges due to two key factors. 
First, it requires cooperation between heterogeneous processors 
(CPU and GPU) with distinct hardware characteristics, across 
UMA platforms that vary in their interconnects and configurations. 
Second, CPU-GPU communication serves different purposes, such as 
data-plane transmission and control-plane coordination, each with 
unique behavioral patterns and requirements. 
To facilitate analysis, we decompose the execution flow of 
CPU-GPU communication via memory sharing into two stages: 
\emph{sending} and \emph{receiving}, 
with unified memory serving as the boundary.
These stages can be implemented independently and combined flexibly, 
enabling systematic analysis of correctness and performance.

\subsection{Correctness of Memory Sharing}
\label{sec:correctness}

Communication via memory sharing involves four operations:  
\begin{itemize}[leftmargin=2.3ex,topsep=0.25ex]
    \item \textbf{\Bind} establishes a shared memory region and configures 
	its attributes separately for the sender and receiver.
    \item \textbf{\Write} populates the memory 
	with data from the sender.
	\item \textbf{\Read} retrieves data from the memory 
	for the receiver. 
	\item \textbf{\Unbind} revokes access permissions and frees memory.
\end{itemize}

\stitle{Methodology.}
We model memory sharing as two stages:
sending (\Bind-\Write-\Unbind) and receiving (\Bind-\Read-\Unbind).
Each stage involves a single processor (CPU or GPU) and can be
configured and implemented independently (e.g., memory attributes).
We empirically test candidate sending and receiving combinations in both CPU--GPU
directions under selected memory configurations, keeping control state separate
from the tested data.
A single-thread persistent GPU kernel participates in each test.
After reading the previous payload, the receiver checks every word of a new
deterministic payload written by the sender.
The verdict is fixed before kernel exit.
An exact match records a successful transfer; stable failure requires
reproducible mismatches after bounded retries.
Timeouts and infrastructure errors are inconclusive, while unsupported
configurations are skipped.

These tests evaluate data visibility between a single sender and receiver.
Write-write contention is outside these tests; atomic
synchronization and contended locks are examined separately
(\S\ref{subsec:acquire-release}).
We assume that the driver and hardware apply the requested page size,
cache policy, and coherence settings.
The results provide empirical evidence for the tested platforms,
configurations, and sharing model, without formal guarantees for general
CPU--GPU shared-memory interactions.
The testing tool is included in the {\sys} artifact.

\subsubsection{{Bind} and {Unbind}}

{\Bind} allocates a shared memory region from unified physical memory 
and configures its memory attributes by setting page tables of 
both the CPU and GPU, which may act as the sender and receiver.
Three critical attributes collectively determine sharing semantics 
and performance characteristics.

\etitle{Page Size.} 
Both CPU and GPU support multiple page sizes. 
The CPU typically uses a small page size (4\,KB) by default.
In contrast, the GPU's page size depends on
the kernel-space driver and user-space runtime implementation. 
While most GPUs support common page sizes, they typically prefer 
a large page size (2\,MB) for better streaming memory access 
performance~\cite{gangulyAdaptivePageMigration2020, johnlinfordDevelopingNVIDIASuperchips2024}. 
When CPU and GPU share memory, the shared region must satisfy the page-alignment
requirements of both the CPU and GPU~\cite{intelcorporationOptimizingMemoryMovement2023}.

\begin{figure}[t]
  \begin{minipage}{1.\linewidth}
    \centering\includegraphics[width=\linewidth,viewport=84 268 327 436,clip=true]{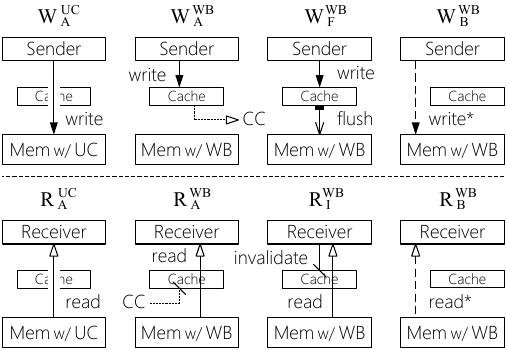}
  \end{minipage} \\[5pt]
  \begin{minipage}{1.\linewidth}
    \caption{\small{\emph{Different implementations of {\Write} and {\Read}.
    }}}
    \label{fig:write-read}
  \end{minipage}  \\[-5pt]
\end{figure}

\etitle{Cache Policy.} 
Write-back (WB) and uncached (UC) are two primary cache policies for memory access. 
UC ensures direct access to shared memory, while WB access data through caches. 
The page table entry (PTE) controls this attribute, affecting only memory accesses
initiated by the associated processor. 
For the CPU, the kernel offers a unified interface to control the cache policy~\cite{pallipadi2008patting}. 
For the GPU, configuring the cache policy requires case-by-case analysis of
the driver's page table configuration code---no general method 
is available~\cite{rossbachOperatingSystemsMust2013a,renHMGExtendingCache2020}. 
GPU accesses on Raptor, Arrow, and Lunar use WB in our evaluation.
On Orin, we use registered host memory, excluding CUDA managed memory.
Registered host memory is uncached in GPU L2 by
default~\cite{nvidiacorporation1CUDATegra2025}\footnote{\footnotesize{On Orin,
registered host memory is uncached in GPU L2 but can be cached in GPU L1.
We bypass GPU L1 for shared-memory loads using the compiler option
``-Xptxas -dlcm=cg''.}}.

\etitle{Data Coherence.} 
Enabling cache coherence (CC) ensures that read operations always retrieve 
the latest version of data and write operations always invalidate stale versions.
This simplifies implementation on the opposite side 
and may enhance performance for memory sharing. 
On our target platforms, 
only the GPU supports controlling CC and enables it by default for memory pages.
The kernel configures this feature by modifying PTE bits. 
Intel's user-space runtimes indirectly expose configurable options 
that allow applications to specify CC strategies. Some Intel platforms 
(e.g., Arrow and Lunar) even allow applications to manipulate 
the relevant PTE bits, 
providing fine-grained control. 

\vspace{1mm}
Page size selection affects only performance (see \S\ref{sec:bind-performance} for details). 
In contrast, cache policy and data coherence determine the correctness of 
memory sharing---specifically, which write and read implementations are needed 
to ensure data is properly written to and read from memory.

\vspace{2mm}
\noindent
{\Unbind} revokes access permissions for both sender and receiver 
to the shared memory region and releases associated resources.
This process reverses the actions performed during {\Bind}. 
Note that unbinding the same memory region twice 
will cause an application crash. 

\begin{table}[t]
\vspace{2mm}
\centering
\begin{minipage}{1.\linewidth}
\caption{\small{\emph{Correct combinations of {\Write}/{\Read} 
and memory attributes on different platforms for CPU-GPU memory sharing.}}}
\label{tab:combinations}
\end{minipage} \\[3pt]
\begin{minipage}{1.\linewidth}
\ra{1.05}
\centering
{\small

\begin{tabular}{|p{0.6cm}||p{2.1cm}|p{4.35cm}|}
\hline

\rowcolor{tablegray}
\textbf{Bind} &
\textbf{Write} &
\textbf{Platform} \\

\hhline{|=::=:=|}
\multirow{1}{*}{UC} & \multirow{1}{*}{Write-Alone}      & CPU (All)~ GPU (Orin) \\ 
\hline
\multirow{3}{*}{WB} & \multirow{1}{*}{Write-Alone}      & CPU (with GPU CC) \\
\cline{2-3}
                    & \multirow{1}{*}{Write-then-Flush} & CPU (All)~ GPU (All) \\
\cline{2-3}
                    & \multirow{1}{*}{Write-Bypass}     & CPU (DMA)~ GPU (Arrow, Lunar) \\
\hline
\end{tabular}

\vspace{1mm}

\begin{tabular}{|p{0.6cm}||p{2.1cm}|p{4.35cm}|}
\hline 

\rowcolor{tablegray}
\textbf{Bind} &
\textbf{Read} &
\textbf{Platform} \\

\hhline{|=::=:=|}
\multirow{1}{*}{UC} & \multirow{1}{*}{Read-Alone}       & CPU (All)~ GPU (Orin) \\ 
\hline
\multirow{3}{*}{WB} & \multirow{1}{*}{Read-Alone}       & CPU (with GPU CC) \\
\cline{2-3}
                    & \multirow{1}{*}{Read-after-Inval} & CPU (All)~ GPU (All) \\
\cline{2-3}
                    & \multirow{1}{*}{Read-Bypass}      & CPU (DMA)~ GPU (Arrow, Lunar) \\
\hline
\end{tabular}

}
\end{minipage} \\[-12pt]
\end{table}

\subsubsection{{Write}}
To ensure data is properly written to memory and visible to the receiver, 
three \Write implementations are available.
Figure~\ref{fig:write-read} illustrates these implementations, and 
Table~\ref{tab:combinations} lists the correct combinations of them and memory attributes.

\etitle{Write-Alone \textup{(\texttt{W${_\text{A}}$})}.} 
A write instruction (e.g., \inst{mov} or \inst{store}) 
can correctly write data to memory only when configured as UC.
However, if the receiver has CC enabled, Write-Alone also functions correctly
with memory configured as WB.

\etitle{Write-then-Flush \textup{(\texttt{W${_\text{F}}$})}.} 
After writes, flushing instructions ensures data is written back to
shared memory, even with WB caching.
Two approaches can evict cached data: 
1) device-specific flush instructions, and 2) external flush commands.
Flush instructions are widely supported on CPUs (e.g., \inst{cl\\flushopt} 
on x86 and \inst{dc} on ARM), although GPU support varies across platforms. 
Arrow and Lunar provide \inst{lsc_fence}~\cite{intelInstructionSetIntel2025}, 
while Raptor and Orin do not include such instructions in official documents 
or compilers.
On our target platforms, the CPU driver can issue external flush commands 
to evict cached data to memory for the GPU via MMIO, 
while the reverse is not supported.
On all three Intel GPUs, the flush command is enqueued as a task
in the hardware queue and processed in order. 
Upon completion of the current kernel, data written by that kernel
is flushed from the GPU's internal cache to memory. In contrast, 
on Orin, a flush command is triggered by writing to a special register (BAR), 
which suggests this process operates independently of kernel execution.

\etitle{Write-Bypass \textup{(\texttt{W${_\text{B}}$})}.} 
Certain specialized instructions enable direct memory writes that 
bypass the private cache hierarchy, 
ensuring correct operation on memory configured as WB. 
For CPUs, non-temporal store instructions (e.g., \inst{movntdq})
write directly to memory and reduce cache pollution. 
However, these instructions serve as hints rather than strict guarantees. 
Furthermore, I/O devices (e.g., SSD) can bypass the cache and write to memory 
directly via DMA. 
For GPUs, store instructions can incorporate cache-bypass modifiers 
to implement Write-Bypass behavior. 
Arrow and Lunar support write instructions that bypass the L2 cache, 
whereas such support was not found on the Raptor and Orin platforms.

\subsubsection{{Read}}

To ensure data from the sender is properly read, 
three \Read implementations are available to retrieve data from memory.
Figure~\ref{fig:write-read} illustrates these implementations, and 
Table~\ref{tab:combinations} lists the correct combinations of them and memory attributes.

\etitle{Read-Alone \textup{(\texttt{R${_\text{A}}$})}.}
A read instruction (e.g., \inst{mov} or \inst{load}) can correctly read data 
from memory only when configured as UC. However, if the sender has CC enabled, 
Read-Alone also functions correctly with memory configured as WB.

\etitle{Read-after-Inval \textup{(\texttt{R${_\text{I}}$})}.}
This implementation first invalidates the private cache, then performs 
read instructions on memory configured as WB. Similar to Write-then-Flush, 
two approaches can invalidate cached data: (1) device-specific invalidation 
instructions, and (2) external invalidation commands. 
Due to data integrity concerns, most platforms do not provide dedicated invalidation 
instructions (or commands). Instead, they rely on flush instructions (or commands)
to implement invalidation, even though no actual writeback occurs. 
Among our target platforms, only the ARM CPU provides the \inst{dc ivac} instruction, 
which invalidates cache lines within a specified region without writeback.

\etitle{Read-Bypass \textup{(\texttt{R${_\text{B}}$})}.}
This implementation mirrors Write-Bypass, 
its counterpart for direct memory writes. 
I/O devices 
use DMA to bypass the cache and read memory directly, 
while GPUs use load instructions with cache-bypass modifiers.

\subsubsection{Acquire and Release}
\label{subsec:acquire-release}

To enable sophisticated CPU-GPU coordination (e.g., locks and barriers), 
control-plane symmetric synchronization relies on two atomic operations
built on memory sharing:

\begin{itemize}[leftmargin=2.3ex,topsep=0.25ex]
  \item \textbf{{\Acquire}} reads the current state to check availability 
  and conditionally writes to establish ownership.
  \item \textbf{{\Release}} writes to the state to relinquish ownership, 
  making it available once more. 
\end{itemize}

\noindent Traditional cross-device implementations of these operations
rely on a \textit{delegation} mechanism~\cite{liManagingScalableDirect2025}, 
where the CPU 
serves as an arbiter to handle acquisition and release requests from the GPU, 
or vice versa. 
For PCIe-connected discrete GPUs, delegation is useful when host
memory lacks CPU--GPU read-modify-write (RMW) atomicity.
System-scope RMW atomicity also depends on 
platform capabilities~\cite{nvidiacorporationAtomicityCUDACore}.
In our tests on NVIDIA discrete GPUs, 
\mbox{\inst{atomicAdd_system}} executed but lost updates under concurrent CPU
increments on mapped host memory.
It produced exact counts without concurrent CPU writers, 
or with them on managed memory.

UMA platforms provide opportunities to optimize symmetric synchronization
by enabling direct contention on shared memory instead of relying on 
delegation to establish a global serialization point. 
Through the unified memory hierarchy, both {\Acquire} and {\Release} 
can be implemented using atomic instructions (e.g., CAS)
directly on shared memory, eliminating the overhead 
and complexity associated with {delegation}.
Our target platforms support such CPU--GPU atomic operations on the
shared memory used by our lock implementations, with CPU mappings
configured as WB.

\subsection{Performance of Memory Sharing}
\label{sec:performance}

\subsubsection{{Memory Binding}}
\label{sec:bind-performance}

\begin{figure}[t]
  \vspace{-1mm}
  \begin{minipage}{1.\linewidth}
    \centering\includegraphics[width=\linewidth]{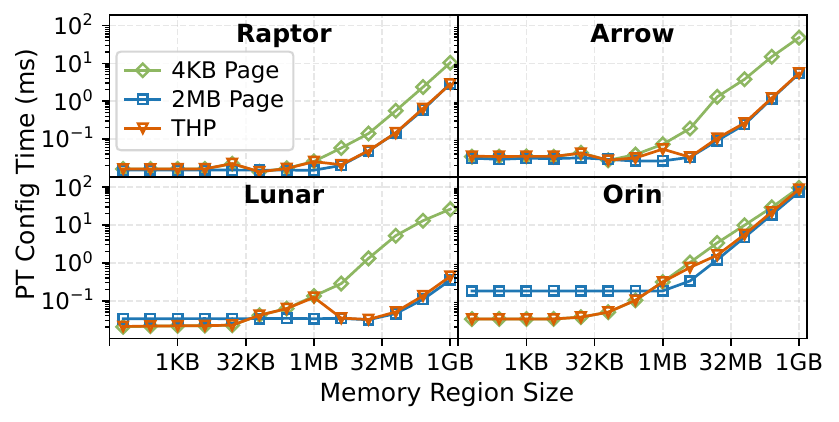}
  \end{minipage} \\[-2pt]
  \begin{minipage}{1.\linewidth}
    \caption{\small{\textit{
	A comparison of page table configuration time for the GPU 
	using different page sizes on our target platforms.
    }}}
    \label{fig:bind}
  \end{minipage}  \\[-5pt]
\end{figure}

Memory binding time represents substantial overhead (about 50\%) 
in scenarios like model loading, where each CPU-GPU communication requires 
one memory binding.
Binding time consists of three parts: memory allocation and page table 
configuration for both the CPU and GPU. Memory allocation time depends 
heavily on memory utilization, while the page table configuration process 
and performance characteristics remain consistent across the CPU and GPU. 
Therefore, we focus on examining page table configuration time for the GPU. 
As shown in Figure~\ref{fig:bind}, page size primarily influences 
memory binding performance and varies across UMA platforms. 
Cache policy and data coherence settings have negligible impact 
since these configurations are applied per PTE.

For large memory regions exceeding 32\,KB, 
the default 4\,KB page size incurs significant time overhead,
dominated by PTE population.
Hugepages (2\,MB) show substantial advantages. 
For small memory regions fitting within a single physical page, 
performance varies across platforms. 
The Intel drivers on Raptor and Arrow show comparable times
for 4\,KB and 2\,MB pages, while 4\,KB pages significantly 
outperform 2\,MB pages on Lunar and Orin (reaching about 5$\times$ on Orin). 
This happens because the driver scans the input memory range 
before modifying page table entries to determine page size and PTE count. 
Registering large pages forces the driver to process a full 2\,MB-aligned region 
even for small memory regions, introducing unnecessary overhead.

\begin{figure*}[t]
  \begin{minipage}{1.\linewidth}
	\hspace{-2mm}
    \centering\includegraphics[width=0.505\linewidth]{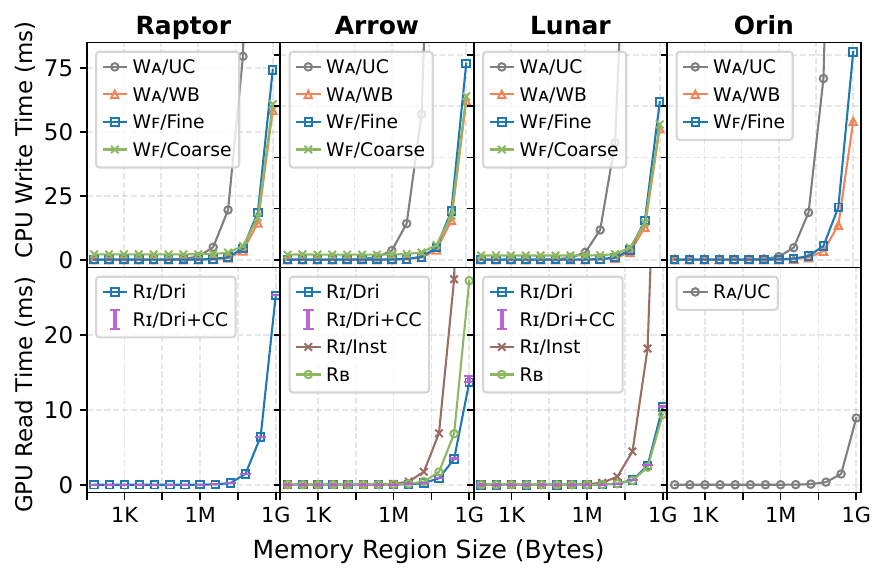}
	\hspace{-2.2mm}
	\centering\includegraphics[width=0.505\linewidth]{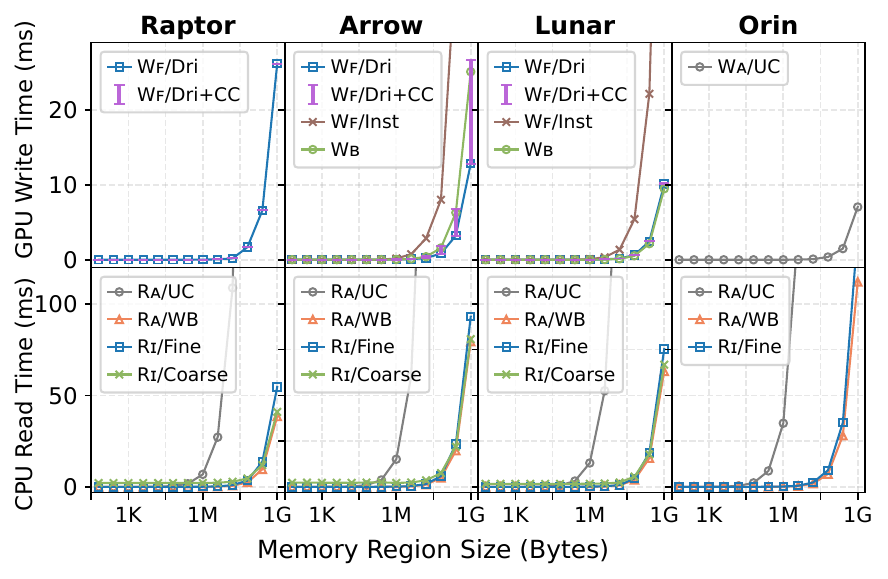}
  \end{minipage} \\[-2pt]
  \begin{minipage}{1.\linewidth}
    \caption{\small{\emph{Performance of {\Write} and {\Read} implementations 
	for memory sharing from CPU to GPU (left) and from GPU to CPU (right). 
  Vertical bars mark the GPU time overhead from enabling CC on Raptor, Arrow, and Lunar.}}}
    \label{fig:per-cpu-gpu}
  \end{minipage}  \\[-5pt]
\end{figure*}

Since 4\,KB pages can incur nearly 100$\times$ higher bind overhead 
and degrade access performance 
due to TLB pressure~\cite{amitIOMMUStrategiesMitigating2010,
jaleelDUCATIHighperformanceAddress2019,zhouImpactPageSize2023}, 
GPU runtimes typically enforce a 2\,MB minimum allocation.
However, this granularity wastes memory for small control-plane messages.
Transparent Huge Pages (THP)~\cite{linuxkerneldevelopmentcommunityTransparentHugepageSupport2025} 
minimizes memory waste 
while automatically promoting memory to hugepages before 
page size becomes a performance bottleneck for memory access.
As shown in Figure~\ref{fig:bind}, THP provides optimal performance in most scenarios.

\stitle{Implications.}
Page size significantly impacts both setup latency and
access performance. While 2\,MB allocation optimizes 
streaming patterns, it proves inefficient for small regions. 
We recommend enabling THP %
and initializing the entire region via \inst{memset} 
to fully populate memory upfront. This approach balances these benefits
against THP drawbacks such as
compaction latency spikes~\cite{panwarMakingHugePages2018,mansiSimPreparingSystem2020}
(Finding~1).

\subsubsection{From CPU to GPU ({CPU$\Rightarrow$GPU})}
\label{sec:cpu2gpu}

Communication from CPU to GPU over UMA involves a CPU-side sending stage 
(CPU \Write) followed by a GPU-side receiving stage (GPU \Read) 
on a shared memory region.
To achieve optimal performance, we select
the most efficient implementation for each stage and combine them.

\stitle{CPU {\Write}.}
Two primary {\Write} implementations are available:
Write-Alone (\texttt{W${_\text{A}}$}) and Write-then-Flush (\texttt{W${_\text{F}}$}).
\texttt{W${_\text{A}}$} is applicable when memory 
is configured as UC %
or as WB with GPU-side CC enabled. %
As shown in Figure~\ref{fig:per-cpu-gpu},
using UC memory (\texttt{W${_\text{A}}$/UC})
forces each write to bypass the cache, increasing write time by up to 300$\times$.
Although hardware write buffers provide 
partial write asynchronicity~\cite{fog2025microarchitecture}, 
their capacity is limited. 
In contrast, WB memory (\texttt{W${_\text{A}}$/WB})
leverages the CPU private cache to enable write combining, 
achieving the lowest write time across all four platforms.

\texttt{W${_\text{F}}$} performance depends on 
the overhead of the mandatory flush operation.
On the ARM-based Orin, 
which is limited to fine-grained flushing \texttt{W${_\text{F}}$/Fine} (\inst{dc civac}),
performance matches \texttt{W${_\text{A}}$/WB} for regions under 16\,MB. 
Beyond this size, flush overhead dominates, increasing latency by up to 50\%.
In contrast, x86 platforms support both user-space fine-grained (\inst{clflushopt}) 
and privileged global flushing (\inst{wbinvd}).
\texttt{W${_\text{F}}$/Coarse} is inefficient for small memory regions 
due to a high fixed overhead ($\sim$2\,ms) from kernel transitions and global flushing.
Conversely, \texttt{W${_\text{F}}$/Fine} is effective up to 256\,MB. 
Beyond this threshold, 
the linear cost of per-line flushing causes approximately 20\% slowdown.
For sufficiently large transfers, the fixed overhead of \texttt{W${_\text{F}}$/Coarse} 
is amortized, allowing performance to approach \texttt{W${_\text{A}}$/WB}.

\begin{figure}[tb]
  \begin{minipage}{1.\linewidth}
    \centering\includegraphics[width=\linewidth]{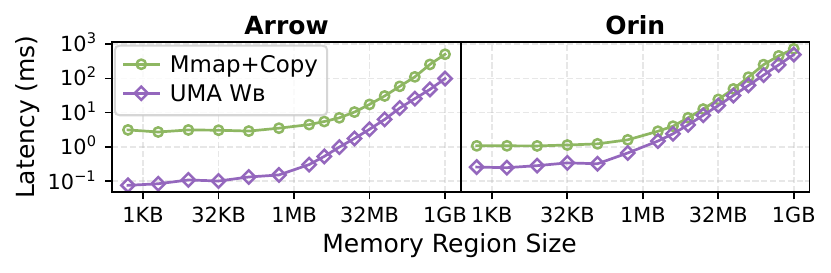}
  \end{minipage} \\[-2pt]
  \begin{minipage}{1.\linewidth}
    \caption{\small{\emph{Performance of CPU {\Write} from SSD to memory.
    }}}
    \label{fig:ssd-load}
  \end{minipage}  \\[-5pt]
\end{figure}

We further study Write-Bypass (\texttt{W${_\text{B}}$}) performance
when data originates from an SSD 
on Arrow (PCIe 5.0) and Orin (PCIe 4.0). 
As shown in Figure~\ref{fig:ssd-load}, 
compared to the mmap-and-copy approach, 
our \texttt{W${_\text{B}}$} implementation eliminates 
overheads of redundant copying, 
page cache management, 
and coherence maintenance. 
In mmap-and-copy, mmap stages data in 
CPU memory before an explicit CPU-to-GPU copy, 
leaving its zero-copy and on-demand semantics 
orthogonal to the SSD-to-GPU transfer evaluated here
and safe to bypass with our DMA path.
This reduces time by 70\% on average
and up to 98\% for 512-byte transfers on Arrow.

\stitle{GPU {\Read}.}
GPU-side implementation performance varies significantly across platforms. 
Therefore, 
we discuss three groups: Intel UHD GPUs (Raptor),
Intel Arc GPUs (Arrow and Lunar), and NVIDIA GPUs (Orin).

\etitle{Intel UHD GPUs.}
Raptor lacks \texttt{LSC}-like instructions for 
memory and cache control. 
It only supports driver-initiated Read-after-Inval (\texttt{R${_\text{I}}$/Dri}), 
which executes at inter-kernel boundaries to perform invalidation. %
Raptor cannot perform intra-kernel invalidation, 
preventing control-plane coordination for applications.
Furthermore, as shown in Figure~\ref{fig:per-cpu-gpu},
enabling CC on Raptor has no discernible impact on GPU {\Read} performance 
(\texttt{R${_\text{I}}$/Dri+CC} vs. \texttt{R${_\text{I}}$/Dri}).
We hypothesize that GPU requests traverse the LLC coherence agents
on Raptor's ring and use its coherence protocol to access the shared LLC.

\etitle{Intel Arc GPUs.}
Arrow and Lunar feature high-performance Intel Arc
GPUs~\cite{intelCoreUltra285HSpecifications, bonenLunarLakeIntel2025a}.
They support fine-grained control over the caching of read/write 
instructions and explicit cache management~\cite{intelcorporationIntelXeGPU2025}.
This offers the most flexible hardware for memory performance evaluation,
supporting Read-after-Inval with both driver-triggered (\texttt{R${_\text{I}}$/Dri})
and instruction-triggered (\texttt{R${_\text{I}}$/Inst}), 
as well as Read-Bypass (\texttt{R${_\text{B}}$}).

\texttt{R${_\text{I}}$/Dri} delivers optimal performance in most cases,
although its use is restricted to inter-kernel boundaries. 
Enabling CC introduces a marginal read overhead on these platforms
due to additional snooping operations required: %
the system must query the snoop filter to locate the latest data 
and broadcast snoop requests to upper-level caches. 
This overhead becomes measurable, ranging 
from 1\% to 6\%, when the memory region size reaches the GB scale.

\texttt{R${_\text{I}}$/Inst} supports both inter-kernel and intra-kernel boundaries.
We implemented \texttt{R${_\text{I}}$/Inst} by performing one flush per block 
with intra-block synchronization. 
This approach balances performance and generality: 
it reduces invalidation frequency compared to per-thread invalidation, 
while eliminating the need for %
implementations to avoid deadlocks 
in global synchronization scenarios. 
However, even with these optimizations,
\texttt{R${_\text{I}}$/Inst} is still the slowest option
on both Arrow and Lunar.

\texttt{R${_\text{B}}$} performance depends heavily on MSC. 
We hypothesize that the MSC coalesces fine-grained \texttt{R${_\text{B}}$}
requests, reducing pressure on the memory controller.
On Lunar (with MSC), \texttt{R${_\text{B}}$} slightly outperforms 
\texttt{R${_\text{I}}$/Dri} for large memory regions
by skipping invalidation operations.
On Arrow (without MSC), \texttt{R${_\text{B}}$} read time
nearly doubles for regions exceeding 16\,MB compared to \texttt{R${_\text{I}}$/Dri}. 
Additionally, \texttt{R${_\text{B}}$} works on both inter-kernel 
and intra-kernel boundaries.

\etitle{NVIDIA GPUs.}
Orin uses UC for registered host memory~\cite{nvidiacorporation1CUDATegra2025},
and its runtime does not expose a way to disable CC.
Consequently, Orin only supports Read-Alone (\texttt{R${_\text{A}}$}) on UC memory.
However, in our streaming microbenchmarks, its performance remains comparable
to that on WB memory; we conjecture that CC overhead is negligible here.
The GPU hides the latency of uncacheable accesses
through extensive thread-level parallelism---a possible rationale behind
NVIDIA's design to use UC policy and enable CC
for registered host-memory access on Orin.
Nevertheless, while GPU workloads generally feature streaming access patterns 
with limited temporal locality~\cite{tianAdaptiveGPUCache2015,
khairyEfficientUtilizationGPGPU2015}, 
real-world applications exhibit this locality at the intra-kernel level. 
Therefore, bypassing the cache incurs a notable performance penalty. 
For example, when the working set fits within the L2 cache, 
UC access experiences a $4\times$ latency increase compared to cached access. 
An L2-optimized GEMM workload on Orin degrades 
by 16\% (from 2,680 to 2,250 GFLOPS) on UC memory.

\stitle{Implications.}
UC memory simplifies sharing but increases CPU write latency by up to
300$\times$ and penalizes GPU workloads with temporal locality.
It should be reserved for cases where neither CC nor access-code
modification is feasible (Finding~2).
Enabling GPU CC permits CPU writes to WB memory without explicit flushing.
Its small GPU-read overhead on Lunar is consistent with low-cost coherence
on MSC-enabled platforms. Arrow also incurs 1--6\% read
overhead without MSC; the larger MSC-related difference arises in GPU writes
(\S\ref{sec:gpu2cpu}) (Finding~3).
For I/O-sourced transfers, DMA into the shared region avoids intermediate
CPU buffering, redundant copies, page-cache management, and explicit
coherence maintenance. This confirms established practice described in
NVIDIA's GPUDirect Storage guidance~\cite{nvidiacorporationGPUDirectStorageOverview}.

We summarize these conclusions as CPU-to-GPU implementation choices in
Table~\ref{tab:suggested-impl}.
For CPU-generated data, use Write-Alone on WB memory
(\texttt{$\text{W}_{\text{A}}^{\text{WB}}$}) with GPU CC enabled;
for I/O data, use DMA-based Write-Bypass
(\texttt{$\text{W}_{\text{B}}^{\text{WB}}$}).
On Intel GPUs, use driver-triggered Read-after-Inval
(\texttt{R$_{\text{I}}^{\text{WB}}$}) between kernels and Read-Bypass
(\texttt{R$_{\text{B}}^{\text{WB}}$}) within kernels on Arrow and Lunar.
Raptor does not support intra-kernel communication.
On Orin, Read-Alone on UC memory (\texttt{R$_{\text{A}}^{\text{UC}}$})
is the only option for registered host memory.

\subsubsection{From GPU to CPU (GPU$\Rightarrow$CPU)}
\label{sec:gpu2cpu}

Communication from GPU to CPU over UMA involves a GPU-side sending stage 
(GPU \Write) followed by a CPU-side receiving stage (CPU \Read) 
on shared memory. 
We follow the same approach to achieve optimal performance.

\stitle{GPU {\Write}.}
As shown in Figure~\ref{fig:per-cpu-gpu}, on Raptor, Lunar, and Orin, 
the performance characteristics of GPU {\Write} implementations 
align with those of GPU {\Read}. %
Both write and read operations saturate memory bandwidth, 
and overheads from flushing or bypassing remain comparable. 
However, on Arrow, GPU write time with CC is $2.07\times$ that without CC
({\texttt{W${_\text{F}}$/Dri+CC} vs. \texttt{W${_\text{F}}$/Dri}}).
The smaller overhead on Lunar is consistent with its MSC reducing
DRAM traffic.

\begin{table}[t]
\vspace{3mm}
\centering
\begin{minipage}{1.\linewidth}
\caption{\small{\emph{Default CPU-GPU memory sharing implementations selected by {\sys}
for our target platforms.
Note that \textup{\texttt{$\text{W}_{\text{A}}^{\text{WB}}$}} 
and \textup{\texttt{$\text{R}_{\text{A}}^{\text{WB}}$}} 
require the opposite GPU to have CC enabled.}}}
\label{tab:suggested-impl}
\end{minipage} \\[3pt]
\begin{minipage}{1.\linewidth}
\ra{1.15}
\centering
{\small
\setlength{\tabcolsep}{4pt}

\begin{tabular}{|c||c|c|c|c|}
\hline 

\rowcolor{tablegray}
\textbf{CPU$\Rightarrow$GPU} &
\textbf{Raptor} &
\textbf{Arrow} &
\textbf{Lunar} &
\textbf{Orin} \\
\hhline{|=:=:=:=:=|}

I/O Device & 
\multicolumn{3}{c|}{\texttt{$\text{W}_{\text{B}}^{\text{WB}}$$\Rightarrow$$\text{R}_{\text{I}}^{\text{WB}}$}} & 
{\texttt{$\text{W}_{\text{B}}^{\text{WB}}$$\Rightarrow$$\text{R}_{\text{A}}^{\text{UC}}$}} \\
\cline{1-5}

Inter-Kernel & 
\multicolumn{3}{c|}{\texttt{$\text{W}_{\text{A}}^{\text{WB}}$$\Rightarrow$$\text{R}_{\text{I}}^{\text{WB}}$}} & 
\multirow{2}{*}{\texttt{$\text{W}_{\text{A}}^{\text{WB}}$$\Rightarrow$$\text{R}_{\text{A}}^{\text{UC}}$}} \\
\cline{1-4}

Intra-Kernel & 
\texttt{N/A} & 
\multicolumn{2}{c|}{\texttt{$\text{W}_{\text{A}}^{\text{WB}}$$\Rightarrow$$\text{R}_{\text{B}}^{\text{WB}}$}}
& 
\\

\hhline{|=:=:=:=:=|}
\rowcolor{tablegray}
\textbf{GPU$\Rightarrow$CPU} &
\textbf{Raptor} &
\textbf{Arrow} &
\textbf{Lunar} &
\textbf{Orin} \\
\hhline{|=:=:=:=:=|}

Inter-Kernel & 
{\texttt{$\text{W}_{\text{F}}^{\text{WB}}$$\Rightarrow$$\text{R}_{\text{A}}^{\text{WB}}$}} &
{\texttt{$\text{W}_{\text{F}}^{\text{WB}}$$\Rightarrow$$\text{R}_{\text{I}}^{\text{WB}}$}} &
{\texttt{$\text{W}_{\text{F}}^{\text{WB}}$$\Rightarrow$$\text{R}_{\text{A}}^{\text{WB}}$}} &
\multirow{2}{*}{\texttt{$\text{W}_{\text{A}}^{\text{UC}}$$\Rightarrow$$\text{R}_{\text{A}}^{\text{WB}}$}} \\
\cline{1-4}

Intra-Kernel & 
\texttt{N/A} & 
{\texttt{$\text{W}_{\text{B}}^{\text{WB}}$$\Rightarrow$$\text{R}_{\text{I}}^{\text{WB}}$}} & 
{\texttt{$\text{W}_{\text{B}}^{\text{WB}}$$\Rightarrow$$\text{R}_{\text{A}}^{\text{WB}}$}} 
& 
\\
\hline
\end{tabular}
}
\end{minipage}  \\[-10pt]
\end{table}

\stitle{CPU {\Read}.}
On the CPU side, read and write performance remains comparable 
across all four platforms. 
Note that on the Arm CPU, the invalidation instruction (\inst{dc ivac})
is much slower than the flush instruction (\inst{dc civac}), 
even though it skips writeback. 
Therefore, on Orin, \texttt{R${_\text{I}}$/Fine} uses 
the flush instruction, consistent with \texttt{W${_\text{F}}$/Fine}.

\stitle{Implications.}
CC simplifies CPU reads with negligible GPU-write overhead on Lunar,
which has an MSC. On Arrow, without an MSC, GPU write time with CC is
2.07$\times$ that without CC, making coherence a significant cost
(Finding~3).

Table~\ref{tab:suggested-impl} summarizes GPU-to-CPU choices.
For inter-kernel communication on Intel platforms, use driver-triggered
Write-then-Flush (\texttt{W$_{\text{F}}^{\text{WB}}$}).
Raptor and Lunar retain GPU CC with CPU Read-Alone
(\texttt{R$_{\text{A}}^{\text{WB}}$}); Arrow disables GPU CC to avoid
coherent-write overhead and uses CPU Read-after-Inval
(\texttt{R$_{\text{I}}^{\text{WB}}$}) to invalidate stale cache lines before reading.
For intra-kernel communication, use GPU Write-Bypass
(\texttt{W$_{\text{B}}^{\text{WB}}$}): Lunar retains CC with CPU Read-Alone,
whereas Arrow disables CC and uses CPU Read-after-Inval
(\texttt{R$_{\text{I}}^{\text{WB}}$}) to avoid coherent-write overhead.
Raptor does not support intra-kernel communication.
On Orin, use GPU Write-Alone on UC memory
(\texttt{W$_{\text{A}}^{\text{UC}}$}) and CPU Read-Alone on WB memory
(\texttt{R$_{\text{A}}^{\text{WB}}$}).

\subsubsection{High Contention}

\begin{figure}[tb]
  \begin{minipage}{1.\linewidth}
    \centering\includegraphics[width=\linewidth]{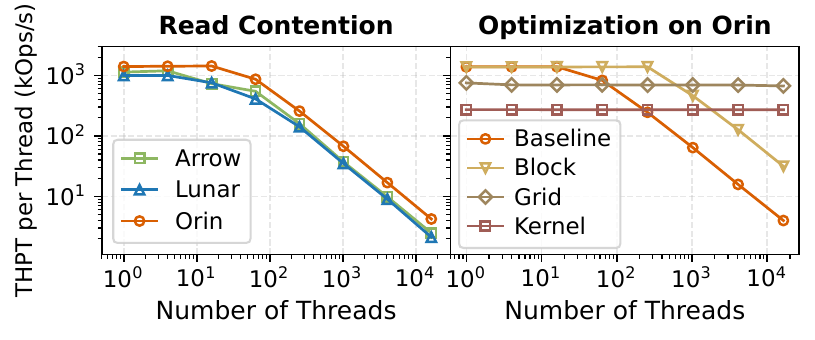}
  \end{minipage} \\[-2pt]
  \begin{minipage}{1.\linewidth}
    \caption{\small{\emph{
        GPU \texttt{R${_\text{B}}$} performance as the thread count increases.
        The number of threads per block is up to 256. %
    }}}
    \label{fig:gpu-contention}
  \end{minipage}  \\[-5pt]
\end{figure}

Control-plane coordination suffers from performance degradation 
due to massive contention on small shared data, 
such as concurrent reads on the GPU side.
As shown in Figure~\ref{fig:gpu-contention} (left), on all platforms,
Read-Bypass (\texttt{R$_{\text{B}}$}) performance degrades sharply 
as thread count increases.
We evaluated three approaches to limit concurrency on Orin 
(see Figure~\ref{fig:gpu-contention} right):

\begin{itemize}[leftmargin=2.3ex,topsep=0.25ex]
  \item \textbf{{Baseline}} uses no barriers---all threads read 
  concurrently.
  \item \textbf{{Block}} enables block-level barriers---only one thread 
  in each block reads the shared data. 
  It offers a simple implementation that scales well for thread 
  counts up to 1,000.	
  \item \textbf{{Grid}} enables grid-level barriers---only one thread 
  across all blocks reads the shared data.
  It requires support for Cooperative Groups~\cite{nvidiacorporationCooperativeGroupsFlexible2017}
  or atomic operations with deadlock avoidance techniques~\cite{guptaStudyPersistentThreads2012}. 
  While incurring a fixed overhead, this approach offers stable performance 
  and outperforms Block when exceeding 1,000 threads.
  \item \textbf{{Kernel}} uses a separate kernel with one thread to read shared data.
  Its performance is consistently lower than Grid due to kernel launch overhead, 
  but it avoids deadlock risks and requires no special instructions.
\end{itemize}

\stitle{Implications.} 
High GPU concurrency
severely degrades performance in control-plane coordination,
making contention management essential. %
We recommend \emph{block-level barriers} to mitigate contention when 
thread counts are below 1,000.
For higher concurrency, \emph{grid-level barriers} are preferred to 
eliminate contention entirely, though they require more complex implementations.
The kernel-level approach should be used only when source code modification is infeasible
(Finding~4).

\section{{libumsh} and Platform Extension}
\label{sec:impl}

Drawing on these findings, 
we built {\sys}, a library that delivers correct and efficient 
CPU-GPU memory sharing across UMA platforms.
Developers compose sending and receiving processes from 
six core operations: \texttt{bind}, \texttt{unbind}, \texttt{write}, \texttt{read}, 
\texttt{acquire}, and \texttt{release}. 
On top of these, {\sys} provides high-level helpers 
for control-plane coordination:
\texttt{wait} and \texttt{set} (built on \texttt{read} and \texttt{write})
and \texttt{lock} and \texttt{unlock} (built on \texttt{acquire} and \texttt{release})
for asymmetric and symmetric synchronization, respectively.

Applications specify the transfer direction, I/O origin, and inter-kernel or
intra-kernel boundary. Given these requirements and the compile-time target
platform, {\sys} automatically selects default {\Read} and {\Write} implementations
and cache and coherence settings (Table~\ref{tab:suggested-impl}).
The backend handles platform-specific binding and cache maintenance.

We deploy {\sys} on all four UMA platforms (see \S\ref{sec:target}). %
It builds on the stock Intel GPU runtime on Raptor 
and the CUDA runtime on Orin;
Arrow and Lunar additionally require custom runtime modifications 
to expose the fine-grained CC control APIs.
The core logic spans about 1,000 lines of code (LoC). 
{\sys} also ships with a companion suite of correctness tests and 
performance benchmarks for characterizing hardware behavior on new platforms.
With compile-time defaults, template-based backend dispatch, and inlined
functions, {\sys} adds less than 1\% overhead over manually applying the same
implementation choices.

\subsection{Porting Workflow}
\label{sec:impl-extension}
Extending {\sys} to a new UMA platform requires two tasks: 
identifying which platform properties behind our findings 
still apply on the target, and 
learning how to configure each through the APIs exposed by 
its driver and runtime. We use the correctness 
and performance suite from \S\ref{sec:sharing} to validate behavior, 
and rely on AI-assisted reading of code and documentation to narrow the search. 
We distill the procedure into a repeatable three-step workflow:

\begin{itemize}[leftmargin=2.3ex,topsep=0.25ex]
    \item \textbf{Locate APIs.}
	 AI-assisted scanning of runtimes, drivers, kernel modules, and 
	 documentation surfaces the APIs that configure page size, cacheability, and coherence,
	 and flags internal behaviors with no public API.
    \item \textbf{Validate behavior.}
	 With AI assistance, we adapt our tests and benchmarks to the identified
	 controls, retaining the sending and receiving model and verdict criteria in
	 \S\ref{sec:correctness}.
    \item \textbf{Adapt the platform backend.}
	 We map the validated controls to {\sys}'s per-platform backend,
	 patching the driver only when a missing control must be exposed.
\end{itemize}

\subsection{DGX Spark Extension}
\label{sec:impl-spark}
This extension tests whether our page-size, cache-policy, and coherence
abstractions support a two-way-coherent UMA platform while preserving
{\sys}'s high-level APIs.

\stitle{Platform details.}
We applied this workflow as a post-study extension to NVIDIA DGX
Spark~\cite{skende171NVIDIAGB102026}.
Its GB10 Grace Blackwell Superchip integrates a 20-core Arm CPU
(10 Cortex-X925 and 10 Cortex-A725 cores), a Blackwell GPU, and
128\,GB of LPDDR5x unified system memory. The 256-bit memory interface
provides 273\,GB/s bandwidth~\cite{nvidiacorporationNVIDIADGXSpark2025}.
The CPU and GPU communicate over NVLink-C2C
with AMBA CHI interoperability, and a 16\,MB MSC
serves both CPU and GPU accesses.
Spark shares Orin's NoC-with-MSC organization, but supports two-way
CPU--GPU coherence, extending beyond the core study's one-way coherence.

\stitle{Library extension.}
Following \S\ref{sec:impl-extension}, AI-assisted inspection of NVIDIA's
open-source GPU kernel modules, UVM driver, CUDA runtime behavior, and
documentation located page-size, cacheability, and two-way CC controls.
Analysis of about 50\,K lines of driver code identified APIs and internal
controls; a roughly 200-line kernel patch exposed controls without public interfaces.
Following our suite, AI generated about 16\,K lines of correctness tests and
performance microbenchmarks. The {\sys} backend required about 100 lines to
map library operations to the validated controls. Measurements and case studies
use the same interfaces and helpers, with no
Spark-specific application changes.
The backend independently selects GPU cacheability (WB or UC) and
coherence (one-way or two-way).

\stitle{Performance of memory sharing.}
Spark preserves THP as a practical default for shared-memory binding.
The CPU-side recommendations also carry over: Write-Alone on WB memory
(\texttt{W$_{\text{A}}^{\text{WB}}$}) for CPU-generated data and
Write-Bypass (\texttt{W$_{\text{B}}^{\text{WB}}$}) for I/O-sourced data.
Our results favor UC for large streaming GPU writes
(\texttt{W$_{\text{A}}^{\text{UC}}$}), while coherent WB remains practical
for GPU reads (\texttt{R$_{\text{A}}^{\text{WB}}$}) and preferable for small
GPU-generated control flags
(\texttt{$\text{W}_{\text{A}}^{\text{WB}}\Rightarrow\text{R}_{\text{A}}^{\text{WB}}$}).
For the largest streaming transfers, GPU Read-Alone and Write-Alone on
WB memory with two-way CC lose about 8\% and up to 27\% throughput,
respectively, relative to their UC counterparts.
In our GPU polling benchmark, \emph{kernel-level} throttling has a
lower fixed overhead than on Orin and is preferred under high contention;
\emph{block-level barriers} remain an alternative when polling
must stay in the original kernel.
\ref{sec:app-spark} provides detailed measurements, implementations,
and case-study results.

\section{Case Study}
\label{sec:apply}

To demonstrate 
our findings and the library, 
this section presents case studies that optimize nine
real-world applications on our four target UMA platforms.
The application integrations use {\sys}'s default platform implementations.
Due to space constraints, 
we show only representative results from these platforms.
Optimized systems are marked as ``+Opt''.
Intermediate variants isolate the effects of individual implementation choices.

\subsection{Data-plane Transmission}
\label{subsec:coarse}

Data-plane transmission typically occurs across kernel boundaries. 
Some scenarios are sensitive to {\Bind} latency 
while others prioritize {\Read} and {\Write} latency. 

\subsubsection{CPU-to-GPU Transfer}

CPU-to-GPU transfer is essential for tasks 
where the CPU prepares data for GPU consumption. 
Common scenarios include model weight loading, 
inference input preparation, texture uploading, 
and software video decoding.

\begin{figure}[tb]
  \begin{minipage}{1.\linewidth}
    \centering\includegraphics[width=\linewidth]{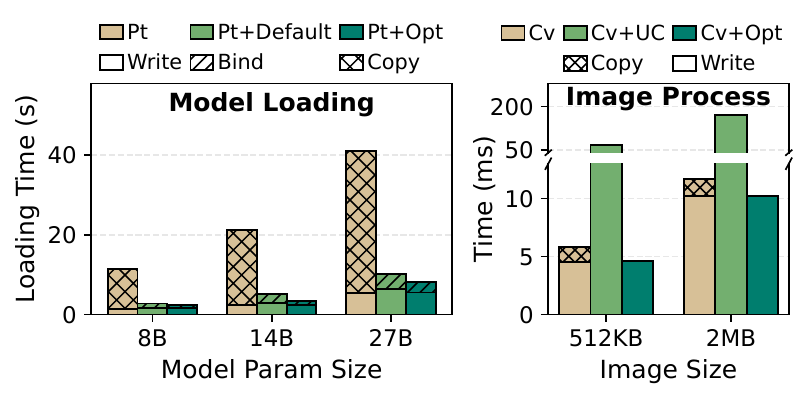}
  \end{minipage} \\[-2pt]
  \begin{minipage}{1.\linewidth}
    \caption{\small{\emph{
        Data-plane optimization from CPU to GPU on Arrow.
    }}}
    \label{fig:case-torch-load}
  \end{minipage}  \\[-8pt]
\end{figure}

\stitle{Model loading.}
PyTorch conventionally loads models by loading (Write) 
and copying them to GPU (Copy). 
We optimize this by loading directly into unified memory 
using \texttt{W${_\text{B}}$}, eliminating the redundant copy. 
Implemented on Arrow ($\sim$300\,LoC), 
we evaluated this across various LLMs 
using both default 4\,KB pages (Pt+Default) and THP (Pt+Opt)
to assess page-size effects on {\Bind} (see Finding~1).
Notably, the baseline copy metric includes lazy {\Bind} overhead.
As shown in Figure~\ref{fig:case-torch-load} (left),
by bypassing the copy phase, 
our default approach reduces loading time by 75\% 
and halves memory consumption. 
THP yields an additional 5\% time reduction 
by minimizing {\Bind} overhead.

\stitle{Image preprocessing.}
Image preprocessing is a critical stage in multimodal inference. 
We use Arrow to compare CPU UC/WB access costs with GPU-side CC (see Finding~2).
We optimized OpenCV~\cite{opencvteamOpenComputerVision2025}
by intercepting its memory allocator to employ unified memory, 
bypassing the copy.
With \sys, the overall modification requires 
only a few LoC.
We evaluated the time of CPU processing (Write) 
and GPU transfer (Copy). 
As shown in Figure~\ref{fig:case-torch-load} (right),
while direct access eliminates copy overhead, 
employing naive UC memory (Cv+UC) increases 
the write time by 20$\times$. 
In contrast, selecting WB memory with GPU-side CC 
(Cv+Opt) bypasses the copy 
step without degrading write performance, 
reducing total time by 16\%.

\subsubsection{GPU-to-CPU Transfer}
GPU-to-CPU transfer is required for 
CPU consumption of GPU outputs. 
Key examples include
getting inference results 
and handling screen captures for recording or streaming.

\begin{figure}[tb]
  \begin{minipage}{1.\linewidth}
    \centering\includegraphics[width=\linewidth]{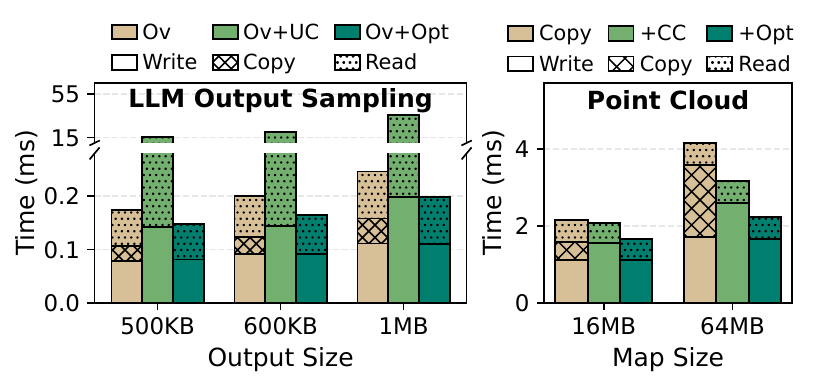}
  \end{minipage} \\[-2pt]
  \begin{minipage}{1.\linewidth}
    \caption{\small{\emph{
        Data-plane optimization from GPU to CPU on Arrow.
    }}}
    \label{fig:case-sample-point}
  \end{minipage}  \\[-5pt]
\end{figure}

\stitle{LLM output sampling.}
During LLM inference with OpenVINO GenAI~\cite{openvinotoolkitOpenVINOGenAI2025}, 
model logits must be transferred to the CPU 
for sampling after each iteration. 
We use Arrow to examine CPU-read cacheability (see Finding~2).
We optimize this by substituting the default output tensor 
with a pre-allocated buffer via {\Bind}, 
bypassing data copying.
This integration requires only $\sim$50\,LoC.
As shown in Figure~\ref{fig:case-sample-point} (left),
using naive UC memory (Ov+UC) incurs a high overhead for {\Read}. 
In contrast, Arrow's default configuration (Ov+Opt) disables GPU CC and pairs
\texttt{W$_{\text{F}}^{\text{WB}}$} with \texttt{R$_{\text{I}}^{\text{WB}}$}:
the CPU invalidates its cache before reading WB memory.
This eliminates copying without degrading {\Read} performance.
It results 
in an 18\% reduction in end-to-end time.

\stitle{Sparse point cloud processing.}
In 3D spatial data processing, 
converting sparse point clouds to dense images 
on the GPU for CPU-side processing~\cite{wuSqueezeSegConvolutionalNeural2018,tuRealTimeStreamingPoint2019} 
often incurs significant transmission costs. 
On Arrow, an initial optimization using
\texttt{W${_\text{F}}$} and \texttt{R${_\text{A}}$} with CC (+CC)
skipped the copy phase, but as shown
in Figure~\ref{fig:case-sample-point} (right),
it incurred a 1.5$\times$ {\Write} time penalty 
because Arrow lacks MSC and the \emph{high-intensity GPU writes}
were bottlenecked by CC (see Finding~3).
For +Opt, {\sys} uses Arrow's default inter-kernel configuration in
Table~\ref{tab:suggested-impl}, with GPU-side CC disabled and
CPU cache invalidation before reading.
This retains zero-copy benefits
while mitigating {\Write} overhead, 
achieving a 34\% reduction in data transfer 
latency compared to the copy baseline.

\subsection{Control-plane Coordination}
\label{subsec:fine}

Control-plane coordination employs small memory regions 
and prioritizes throughput for the {\Read} and {\Write} operations.
There is no significant difference in latency 
among the various {\Read} and {\Write} implementations. 

\subsubsection{Asymmetric Synchronization}

Asymmetric synchronization is a basic pattern 
for sequencing cross-device operations and enforcing dependencies.
Common scenarios include the CPU waiting for GPU RPC calls or
the GPU awaiting host data preparation. 
Traditional approaches rely on 
high-overhead mechanisms, such as explicit kernel launches 
or runtime events. 
UMA platforms allow us to leverage \textit{direct access} 
for lightweight control.
This mechanism employs a one-way signal that repeatedly reads a flag and writes
it once, using \texttt{wait}/\texttt{set} in \sys.

\begin{figure}[tb]
  \begin{minipage}{1.\linewidth}
    \centering\includegraphics[width=\linewidth]{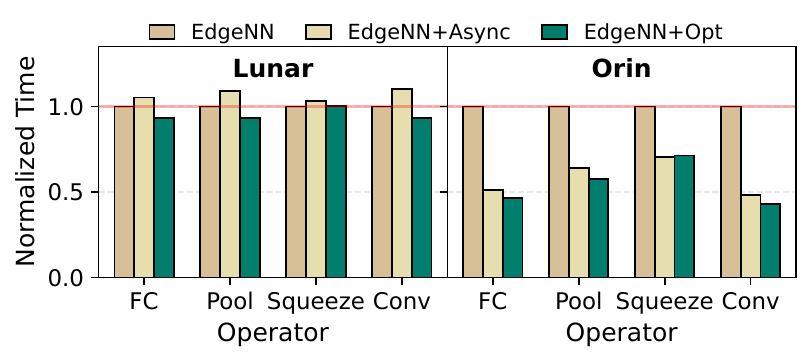}
  \end{minipage} \\[-2pt]
  \begin{minipage}{1.\linewidth}
    \caption{\small{\emph{
        Control-plane optimization for hybrid inference on Lunar (left) and Orin (right).
    }}}
    \label{fig:case-hetero}
  \end{minipage}  \\[-5pt]
\end{figure}

\stitle{Hybrid inference.} 
To address the synchronization overhead
in CPU-GPU hybrid inference, 
we leveraged EdgeNN~\cite{zhangEdgeNNEfficientNeural2023a}
to evaluate a control flow optimization 
based on direct access signaling.
We use Lunar and Orin to compare synchronization across two vendors'
NoC-with-MSC platforms.
By default, EdgeNN employs a serialized execution model: 
launching the GPU kernel, performing CPU computation, 
and then blocking until GPU completion.
To mitigate the high control overhead associated 
with frequent kernel launches, 
we initially decoupled the launch and compute phases 
using native runtime \textit{events} (EdgeNN+Async),
with $\sim$100\,LoC.
However, as shown in Figure~\ref{fig:case-hetero}, 
although this approach prevents the 
kernel launch from blocking CPU execution, 
the runtime overhead proved detrimental 
on the Lunar platform, increasing latency to
as much as $1.1\times$ the baseline.
We replaced the \textit{events} with 
our \texttt{wait}/\texttt{set} operations ($\sim$50 LoC),
enabling UMA-based lightweight signaling (EdgeNN+Opt).
This optimization finally reduced operator 
execution time by 4.9\% on Lunar and 45.4\% on Orin on average
compared to EdgeNN.

\begin{figure}[tb]
  \begin{minipage}{1.\linewidth}
    \centering\includegraphics[width=\linewidth]{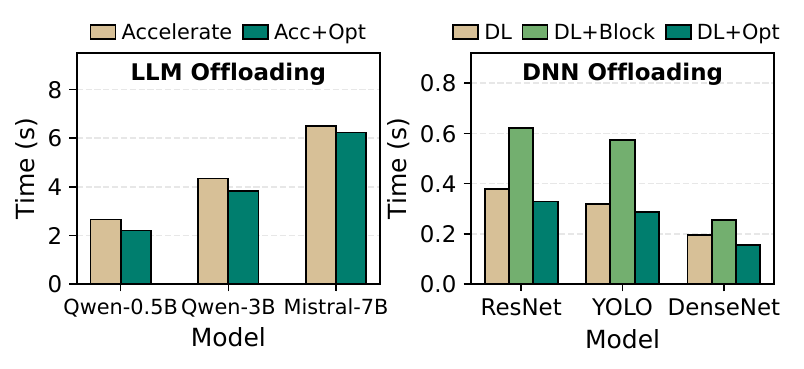}
  \end{minipage} \\[-2pt]
  \begin{minipage}{1.\linewidth}
    \caption{\small{\emph{Offloading inference: Accelerate on Arrow (left)
        and Demand Layering on Orin (right).
    }}}
    \label{fig:case-offload-sync}
  \end{minipage}  \\[-5pt]
\end{figure}

\stitle{Model offloading.}
We address the synchronization overhead between weight loading 
and kernel execution. 
First, we targeted 
Accelerate~\cite{HuggingfaceAccelerate2025} on Arrow
to demonstrate a black-box scenario
where JIT-generated operators preclude source code modification.
We decoupled the weight-loading and kernel-launch phases 
with a \emph{kernel-level approach} to control the subsequent 
kernel execution, managing dependencies via 
\sys~($\sim$200\,LoC, Acc+Opt).
As shown in Figure~\ref{fig:case-offload-sync} (left),
evaluation on a 10-token prefill workload 
demonstrates a TTFT reduction of 11\%.

To apply Orin's contention-control techniques (Figure~\ref{fig:gpu-contention}),
we optimized Demand Layering~\cite{jiDemandLayeringRealTime2022b} on Orin,
where operator source code can be modified directly.
Modifications to individual operators 
required only a few LoC, 
while the remaining integration 
involved $\sim$200\,LoC.
Given that concurrent $\texttt{R}_B$ operations 
heavily impact read throughput, 
we evaluated the use of both \emph{block-} and 
\emph{grid-level barriers} to throttle concurrency. 
As illustrated in Figure~\ref{fig:case-offload-sync} (right),
results indicate that  
\emph{block-level barriers} (DL+Block) backfired, increasing
execution time to $1.8\times$ the baseline.
Conversely, while \emph{grid-level barriers} (DL+Opt) introduced 
complexity due to deadlock avoidance, 
this effectively managed contention, reducing execution time by 14\%.

\begin{figure}[t]
  \begin{minipage}{1.\linewidth}
    \centering\includegraphics[width=\linewidth]{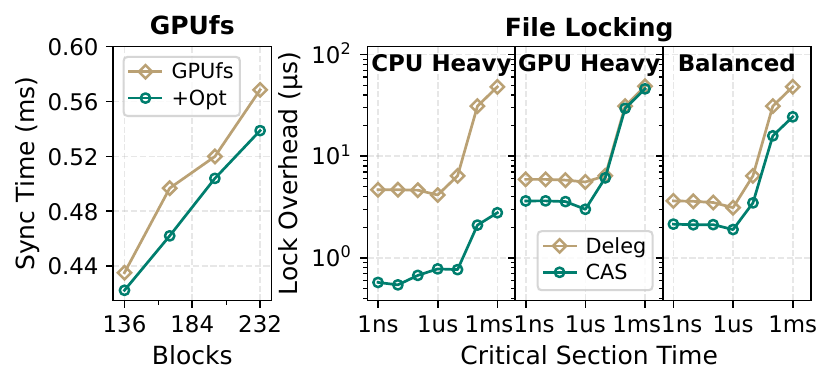}
  \end{minipage} \\[-0pt]
  \begin{minipage}{1.\linewidth}
    \caption{\small{\emph{
        Control-plane optimization for GPUfs and file locking on Orin.
        Deleg denotes Delegation on the right figure.
    }}}
    \label{fig:case-fs}
  \end{minipage}  \\[-5pt]
\end{figure}

\stitle{GPU file access.}
Beyond computation and memory, invoking system services 
constitutes a critical workload. 
We use Orin to compare block- and grid-level coordination under
the polling contention identified in Figure~\ref{fig:gpu-contention}.
GPUfs~\cite{silberstein2014gpufs} facilitates this 
by delegating GPU file requests to the CPU 
via memory-based synchronization with \emph{block-level barriers} (GPUfs).
However, our analysis reveals that for kernel-wide file operations, 
this approach still suffers from contention.
At high block counts, \emph{grid-level barriers} (+Opt) reduce synchronization time
by about 7\% (Figure~\ref{fig:case-fs}, left).

\subsubsection{Symmetric Synchronization}
\label{subsec:symmetric}

Symmetric synchronization is characterized by 
the CPU and GPU contending for shared state, 
necessitating atomic operations to prevent race conditions. 
Examples include mediating 
concurrent writes in a shared filesystem or 
updating arrival counters during barrier synchronization. 
We select the cross-device lock as the representative case
for analysis.

\stitle{Cross-device file locking.} 
GoFS~\cite{liManagingScalableDirect2025} manages shared file consistency 
between the CPU and GPU via a GPU-resident delegation server. 
To reduce the overhead of this centralized approach, 
we implemented a direct-access \textit{CAS lock} 
(via \texttt{lock} and \texttt{unlock} in \sys)
on Orin, which supports CPU--GPU atomics on shared host memory,
to replace the default \textit{delegation lock}.
We evaluated the lock overhead, defined as the time 
per lock operation minus the critical-section time,
under low contention where one CPU thread
and one GPU thread compete for the same lock.
As shown in Figure~\ref{fig:case-fs} (right), 
in the CPU-heavy scenario, 
for file operations whose critical-section time
is on the microsecond scale,
the \textit{CAS lock} reduces lock overhead
by about 90\% compared with the \textit{delegation lock}.
This advantage arises because \textit{CAS} operates 
directly on shared memory, avoiding the RPC overhead. 

\begin{figure}[t]
  \vspace{1mm}
  \begin{minipage}{1.\linewidth}
    \centering\includegraphics[width=\linewidth]{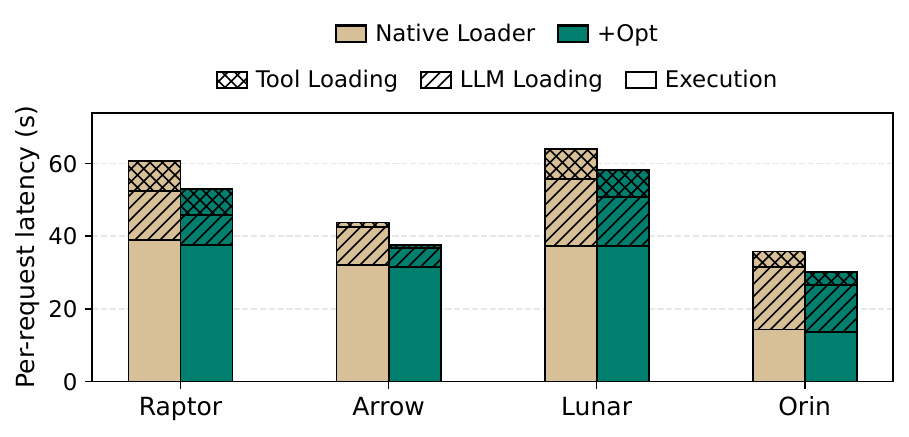}
  \end{minipage} \\[-2pt]
  \begin{minipage}{1.\linewidth}
    \caption{\small\emph{Estimated per-request agent-serving latency on
    TaskBench traces, including three tool calls.}}
    \label{fig:case-end-to-end}
  \end{minipage}  \\[-5pt]
\end{figure}

\subsection{End-to-End Evaluation}
\label{subsec:end-to-end}

Our agent-serving evaluation uses TaskBench
traces~\cite{shenTaskBenchBenchmarkingLarge2024}, in which a GLM-4.7-Flash
orchestrator invokes ViT, SD-Turbo, and DETR sequentially.
The LLM runs in \texttt{llama.cpp} and the tools in PyTorch;
runtimes remain warm while models reload on demand.
Across four platforms, we compare {\sys} with native loaders using estimated
per-request latency: the sum of the separately measured means of orchestration
time and the three tools' combined execution time, both including model loading.

Figure~\ref{fig:case-end-to-end} shows that {\sys} reduces estimated
per-request latency by 9.0--15.9\%, with model loading accounting for
most of the savings.
On Arrow, latency decreases by 14.0\% (from 43.78\,s to 37.65\,s).
LLM loading before prefill falls by 51.3\% (from 10.40\,s to 5.06\,s),
accounting for 5.34\,s of the 6.13\,s total saving.
These results show how the model-loading benefits in
Figure~\ref{fig:case-torch-load} translate into lower serving latency.

\section{Discussion}
\label{sec:discussion}

\nostitle{Limited visibility and control through OS abstractions.} 
OS abstractions simplify development
but hide critical parameters (page size, device CC status, 
and cache policy) from user space.
For example, THP treats hugepage allocation as a best-effort hint
and silently falls back to 4\,KB pages under fragmentation, causing
unpredictable performance drops~\cite{kwonIngensHugePage2017,panwarMakingHugePages2018}.
{\sys} mitigates the worst cases by prefaulting THP regions (\S\ref{sec:bind-performance}) 
and patching the driver to expose hidden parameters (\S\ref{sec:impl}), 
but the underlying lack of visibility 
remains a systemic obstacle to correct and efficient CPU-GPU memory sharing.

\stitle{The evolution and caveats of hardware coherence.}
Hardware CC reduces the software burden of memory sharing, 
but using it correctly still demands platform-specific analysis. 
The shift from one-way to two-way coherence 
appears to simplify development, yet in practice it complicates analysis and optimization.
AMD's two-way coherence omits the GPU L0/L1 caches
and still requires Write-Bypass instructions for correctness~\cite{llvmUserGuideAMDGPU2025};
NVIDIA explicitly recommends benchmarking 
because of potential overheads~\cite{nvidiacorporation1CUDATegra2025}.
The DGX Spark extension in \S\ref{sec:impl-spark} illustrates why a wider
coherence domain still requires validation of cache coverage and performance.

\stitle{Beyond CPU-GPU memory sharing.}
Our methodology applies to
other heterogeneous components in UMA platforms, 
such as NPUs and ASICs.
While the logic for analyzing correctness and performance is universal, 
most ISAs and runtimes remain closed and 
undocumented~\cite{chenTVMAutomatedEndtoEnd2018,huFirstLookOndevice2023,kaladeNPUEvalOptimizingNPU2025}. 
The AI-assisted porting workflow (\S\ref{sec:impl-extension}) narrows the
search for controls, as the DGX Spark extension demonstrates. Applying it to
other accelerator classes still requires access to their hardware controls
and validation of cache visibility and synchronization behavior.

\section{Related Work}
\label{sec:relwrk}

\nostitle{Memory performance of UMA platforms.} 
Prior research on UMA memory performance can be categorized into three categories. 
First, many studies examine isolated architectural features, such as 
interconnects~\cite{smithRealizingAMDExascale2024, ruzhanskaia2024rethinking, 
duttaLeakyBuddiesCrossComponent2021a}, 
memory access~\cite{pichaiArchitecturalSupportAddress2014, 
ausavarungnirunMosaicGPUMemory2017, kouAttackDirectoriesARM2022a, 
xuEXAMExploitingExclusive2025}, 
and cache coherence~\cite{powerHeterogeneousSystemCoherence2013, 
alsopSpandexFlexibleInterface2018, 
natarajaEnhancedSystemLevelCoherence2024, 
zuckermanCohmeleonLearningBasedOrchestration2021, 
agarwalSelectiveGPUCaches2016, alsopCaseFinegrainCoherence2022}. 
Second, most are designed for specific platforms, 
limiting their applicability across different architectures~\cite{halawaNVIDIAJetsonPlatform2017, 
hubnerAppleVsOranges2025, geraPerformanceCharacterisationSimulation2018, 
jinPopulationCountIntel2020,damschenCoSchedulingFusedCPUGPU2018,wahlgren2025dissecting}. 
Third, while some cross-platform analyses
exist~\cite{zhangIMLBenchMachineLearning2021, 
cavicchioliMemoryInterferenceCharacterization2017, lumppFrameworkOptimizingCPUiGPU2021,
udkarUniformMemoryArchitecture2025,taUnderstandingImpactFineGrained2017}, 
they either focus on performance data without addressing correctness,
or they target narrow application scenarios.
Our analysis provides a comprehensive methodology 
that spans both correctness and performance across diverse platforms and use cases.

\stitle{CPU-GPU communication.} 
Prior research has extensively investigated and optimized CPU-GPU communication. 
For data-plane transmission, 
researchers have used unified memory to 
achieve transparent data transfer between the CPU and 
GPU~\cite{jungOverlappingHosttodeviceCopy2020,
allenIndepthAnalysesUnified2021,fumeroUnifiedSharedMemory2023}, 
or optimize transmission for specific 
scenarios~\cite{seilerAutomaticGPUData2020,liuExploringQueryProcessing2022}. 
For control-plane coordination, 
optimization efforts have mainly focused on
hybrid computing~\cite{zhangEdgeNNEfficientNeural2023a, 
kimMLayerLowLatency2019, xue2024powerinfer, chenCharacterizingMobileSoC2025, 
xuFastOndeviceLLM2025,liHydrogenContentionAwareHybrid2024} and 
task scheduling~\cite{shen2025xsched,dagliSharedMemorycontentionawareConcurrent2024}. 
Cross-platform libraries such as Kokkos~\cite{kokkosMemorySpaces}
abstract memory spaces, while CUDA~\cite{nvidiacorporation1CUDATegra2025}
provides vendor-specific allocation and registration APIs.
However, these efforts are largely conducted in isolation. 
Our work provides a comprehensive analysis of the application-level 
requirements governing CPU-GPU communication 
and a library that automatically selects implementations across platforms.

\section{Conclusion}
\label{sec:concl}

This paper systematically characterizes the correctness and performance of
CPU-GPU memory sharing on unified memory architectures. By decomposing sharing
into memory binding and paired write and read operations, our methodology exposes
how page size, cache policy, coherence, MSC, and contention
interact across four representative platforms. Our correctness tests and four
performance findings show why direct access alone does not ensure correct and
efficient sharing: implementation choices must account for both the platform's memory hierarchy and the
visibility and synchronization requirements of each sharing pattern.

Guided by this, we developed {\sys}, which encapsulates
platform-specific implementations behind high-level sender-receiver operations
and automatically selects defaults according to application requirements.
Its application to nine real-world systems demonstrates benefits for both
data-plane transmission and control-plane coordination.
The NVIDIA DGX Spark extension applies the same methodology and high-level
APIs to a platform with two-way coherence.
Together, the methodology, validation suite, and library provide a reusable
approach to supporting new UMA platforms through correctness testing,
performance measurement, and adaptation of platform-specific implementations.

\begin{acks}
We sincerely thank our shepherd and the anonymous reviewers 
for their constructive feedback and suggestions. 
This work was supported in part by 
the Fundamental and Interdisciplinary Disciplines Breakthrough Plan 
of the Ministry of Education of China (JYB2025XDXM122)
and the National Natural Science Foundation of China (No. 62672297, 62432010).
\end{acks}

\bibliographystyle{ACM-Reference-Format}
\bibliography{bib/umxpu}

\clearpage
\appendix
\counterwithin{figure}{section}
\counterwithin{table}{section}
\renewcommand{\thefigure}{\Alph{section}.\arabic{figure}}
\renewcommand{\thetable}{\Alph{section}.\arabic{table}}

\renewcommand{\thesection}{Appendix \Alph{section}}

\section{Artifact Appendix}
\label{sec:app-artifact}

This artifact provides the source code of {\sys}, correctness tests,
and microbenchmarks for the EuroSys 2027 paper
``Characterizing CPU-GPU Memory Sharing over Unified Memory Architecture''
by J.~Yang, Z.~Wang, R.~Chen, and H.~Chen.
{\sys} is a library that enables CPU-GPU memory sharing on unified memory
architectures through its data-plane and control-plane implementations.
We provide instructions for building the library and running the tests
and microbenchmarks, along with reference measurements and plotting scripts
for the performance characterization in \S\ref{sec:sharing}.
Our artifact received the ``Artifacts Available'' and
``Artifacts Evaluated (Functional)'' badges from the EuroSys 2027
Artifact Evaluation process. The DOI of our artifact is
\url{https://doi.org/10.5281/zenodo.22909405}.

\stitle{Artifact repository.}
The {\sys} source code and instructions for building and evaluating it on
supported Intel and NVIDIA UMA platforms are available in the following
Git repository: \url{https://github.com/XpuOS/libumsh-artifacts}.

\section{NVIDIA DGX Spark Extension}
\label{sec:app-spark}
\suppressfloats[t]

This appendix reports memory-sharing performance measurements and case studies
for NVIDIA DGX Spark. The platform details, library extension, and main
conclusions are presented in \S\ref{sec:impl-spark}.
All measurements below exercise the Spark backend added to {\sys}.

\begin{figure}[tb]
  \vspace{-2mm}
  \begin{minipage}{1.\linewidth}
    \centering\includegraphics[width=0.73\linewidth]{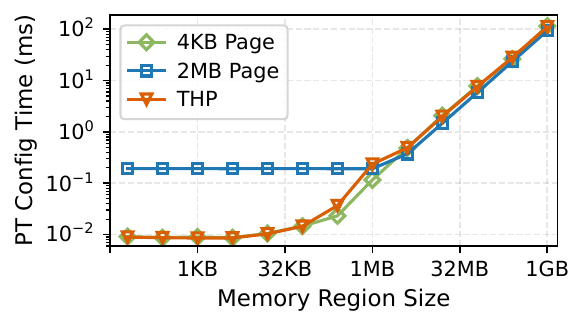}
  \end{minipage} \\[2pt]
  \begin{minipage}{1.\linewidth}
    \caption{\small{\textit{
	Page table configuration time for the GPU
	using different page sizes on Spark.
    }}}
    \label{fig:app-spark-bind}
  \end{minipage}
\end{figure}

\subsection{Performance of Memory Sharing}

This subsection extends the memory-sharing performance study in
\S\ref{sec:performance} to Spark. 
We run the corresponding 
correctness tests and 
performance benchmarks in {\sys} 
to characterize Spark's hardware behavior.
We analyze the same implementation choices for binding,
CPU-to-GPU transfer, GPU-to-CPU transfer, and GPU-side contention. 

\stitle{\Bind.} Figure~\ref{fig:app-spark-bind} shows 
the time to set up a memory region for sharing,
which includes page table configuration and
any associated cache management.
Small regions favor 4\,KB pages or 
THP because 2\,MB mappings impose a 
fixed setup cost, while large streaming 
regions amortize the larger mapping setup. 
Thus, on Spark, THP remains a practical default: it avoids the
fixed setup cost of explicit 2\,MB mappings for small regions while
not introducing a penalty for large regions.

\begin{figure}[tb]
  \vspace{2mm}
  \begin{minipage}{1.\linewidth}
    \centering\includegraphics[width=1\linewidth]{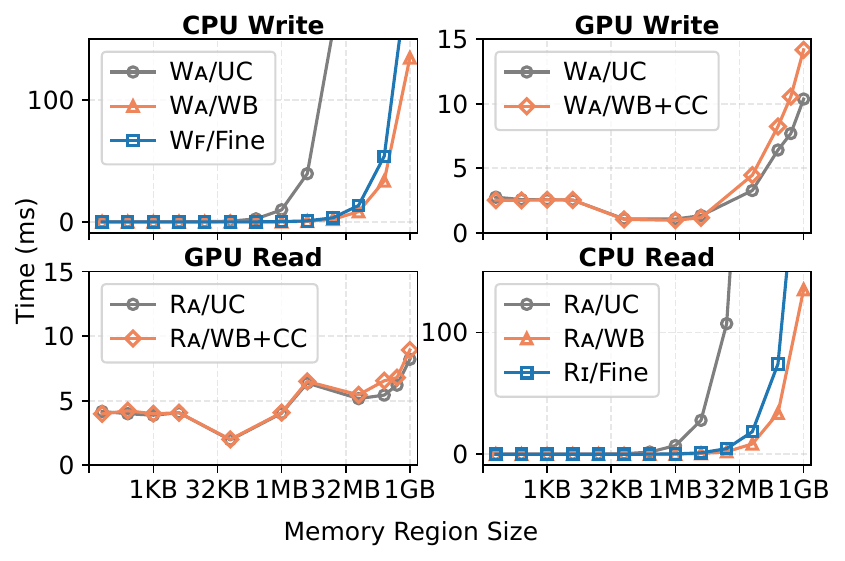}
  \end{minipage} \\[2pt]
  \begin{minipage}{1.\linewidth}
    \caption{\small{\textit{
	Performance of {\Write} and {\Read} implementations for memory sharing on Spark, 
  broken down by CPU and GPU access paths.
  CC denotes two-way coherence.}}}
    \label{fig:app-spark-transfer}
  \end{minipage}  \\[-4pt]
\end{figure}

\stitle{\Write and \Read.}
Figure~\ref{fig:app-spark-transfer} reports the performance of the Spark implementations of
{\Write} and {\Read}. On the CPU side, Spark exposes the standard CPU access
paths used in the main study: \texttt{W${_\text{A}}$/WB} for CPU-generated
data, \texttt{W${_\text{F}}$/Fine} when explicit cache flushing is required,
\texttt{R${_\text{A}}$/WB} for cached reads, and \texttt{R${_\text{I}}$/Fine}
when invalidation is needed before reading. The GPU side exposes a richer set
of choices than Orin because Spark's GPU driver is
open-source: our backend extension can expose additional APIs to {\sys},
including APIs for selecting WB memory and toggling two-way CC support. 

The CPU-side results preserve the main-study choices: WB memory keeps
\texttt{W${_\text{A}}$} efficient for CPU-generated data, while explicit
fine-grained flushing or invalidation is only needed when software must
maintain visibility without relying on coherence. The GPU-side results are
different: hardware-managed two-way CC does not universally replace
software-managed accesses. In our streaming benchmarks, once the shared buffer reaches the MB scale,
enabling two-way CC on WB memory can be slower than the UC
paths. For the largest transfers, \texttt{R${_\text{A}}$/WB+CC} loses about
8\% GPU {\Read} throughput relative to \texttt{R${_\text{A}}$/UC}, while
\texttt{W${_\text{A}}$/WB+CC} loses up to 27\% GPU {\Write} throughput
relative to \texttt{W${_\text{A}}$/UC}. Thus, Spark's two-way CC is useful
for coherent GPU reads and small control flags, but MB-scale streaming GPU writes should
prefer the UC path.

\begin{figure}[tb]
  \vspace{2mm}
  \begin{minipage}{1.\linewidth}
    \centering\includegraphics[width=0.73\linewidth]{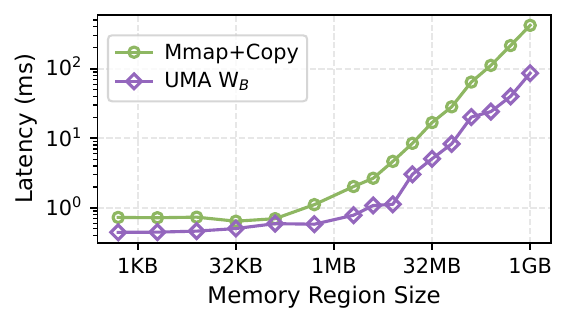}
  \end{minipage} \\[2pt]
  \begin{minipage}{1.\linewidth}
    \caption{\small{\textit{
	   Performance of CPU {\Write} from SSD to memory on Spark. \texttt{W${_\text{B}}$} reduces end-to-end
     loading time relative to the mmap-and-copy path by avoiding extra staging and copy
     overhead.
    }}}
    \label{fig:app-spark-ssd-load}
  \end{minipage}
\end{figure}

Figure~\ref{fig:app-spark-ssd-load} presents 
the performance of Write-Bypass (\texttt{W${_\text{B}}$})
on Spark. For SSD-backed CPU-to-GPU data-plane transmission,
\texttt{W${_\text{B}}$} avoids page-cache staging and an extra memory copy,
so it outperforms the traditional mmap-and-copy path.
Overall, Spark does not change the high-level implementation guidance:
use bypass-style CPU {\Write} for I/O-sourced data.

\begin{table}[tb]
\vspace{4mm}
\centering
\begin{minipage}{1.\linewidth}
\caption{\small{\emph{Recommended CPU-GPU memory sharing implementations on Spark.
Note that \textup{\texttt{$\text{W}_{\text{A}}^{\text{WB}}$}} 
and \textup{\texttt{$\text{R}_{\text{A}}^{\text{WB}}$}} 
rely on CC support.
}}}
\label{tab:app-spark-suggested-impl}
\end{minipage} \\[5pt]
\begin{minipage}{1.\linewidth}
\centering
\small
\setlength{\tabcolsep}{2pt}
\renewcommand{\arraystretch}{1.15}
\begin{tabular}{|>{\centering\arraybackslash}p{2.2cm}||>{\centering\arraybackslash}p{2.2cm}|>{\centering\arraybackslash}p{3cm}|}
\hline
\rowcolor{gray!20}
\textbf{Stage} & \textbf{Case} & \textbf{Recommended path} \\
\hline
{\Bind} & General & THP \\
\hline
\multirow{3}{*}{CPU$\Rightarrow$GPU} & I/O Device &
\texttt{$\text{W}_{\text{B}}^{\text{WB}}\Rightarrow\text{R}_{\text{A}}^{\text{WB}}$} \\
\cline{2-3} & Inter-Kernel &
\multirow{2}{*}{\texttt{$\text{W}_{\text{A}}^{\text{WB}}\Rightarrow\text{R}_{\text{A}}^{\text{WB}}$}} \\
\cline{2-2} & Intra-Kernel & \\
\hline
\multirow{2}{*}{GPU$\Rightarrow$CPU} & Inter-Kernel &
\texttt{$\text{W}_{\text{A}}^{\text{UC}}\Rightarrow\text{R}_{\text{A}}^{\text{WB}}$} \\
\cline{2-3} & Intra-Kernel &
\texttt{$\text{W}_{\text{A}}^{\text{WB}}\Rightarrow\text{R}_{\text{A}}^{\text{WB}}$} \\
\hline
\end{tabular}
\end{minipage}  \\[-2pt]
\end{table}

\stitle{Implications.}
Table~\ref{tab:app-spark-suggested-impl} summarizes the recommended
CPU-GPU memory sharing implementations on Spark. Spark mostly preserves the
main study's recommendations for binding and CPU-side accesses, but changes the
GPU-side recommendation: coherent WB memory remains practical for GPU {\Read} and for
small GPU-generated flags, whereas GPU {\Write} should prefer the
non-coherent UC path for MB-scale streaming transfers.
The UC recommendation applies to streaming transfers; applications with
intra-kernel reuse also need to consider cache locality (\S\ref{sec:cpu2gpu}).

For {\Bind}, THP is recommended for general shared regions because it avoids
the fixed setup cost of explicit 2\,MB mappings for small regions while
preserving large-region performance. 
For CPU-to-GPU communication, DMA-based Write-Bypass
(\texttt{W$_{\text{B}}^{\text{WB}}$}) followed by GPU Read-Alone on WB memory
(\texttt{R$_{\text{A}}^{\text{WB}}$}) is recommended for data from I/O devices
(e.g., SSD), whereas Write-Alone on WB memory
(\texttt{W$_{\text{A}}^{\text{WB}}$}) followed by the same GPU read path is
appropriate for CPU-generated data. In both cases, Spark should keep the GPU
consumer on coherent WB reads rather than switching to UC memory.
For GPU-to-CPU communication, Spark should use the non-coherent UC GPU write
path (\texttt{W$_{\text{A}}^{\text{UC}}$}) for inter-kernel streaming transfers
because it avoids the two-way CC overhead observed for large WB GPU writes. For
intra-kernel flag passing, the coherent WB path
(\texttt{$\text{W}_{\text{A}}^{\text{WB}}\Rightarrow\text{R}_{\text{A}}^{\text{WB}}$})
is preferable because the performance cost of two-way CC is negligible at this scale,
and the CPU can read the flag directly.

\begin{figure}[tb]
  \vspace{2mm}
  \begin{minipage}{1.\linewidth}
    \centering\includegraphics[width=\linewidth]{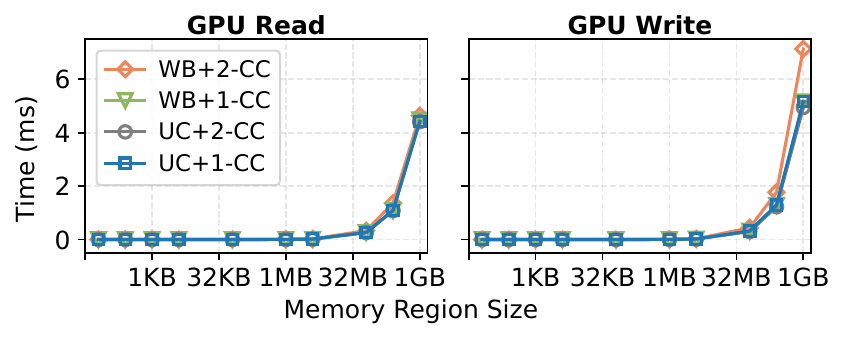}
  \end{minipage} \\[2pt]
  \begin{minipage}{1.\linewidth}
    \caption{\small{\textit{
    GPU {\Read} and {\Write} performance on Spark under 
    different cache and CC policies. 
    1-CC denotes one-way coherence, and 2-CC denotes two-way coherence.
    }}}
    \label{fig:app-spark-gpu-cc}
  \end{minipage}
\end{figure}

\stitle{Two-way coherence.}
To isolate the source of Spark's two-way CC behavior, we further tested GPU
access performance under a $2\times2$ matrix of memory configurations. Spark allows the
backend to configure both cache policy (WB or UC) and CC policy (one-way or
two-way), so Figure~\ref{fig:app-spark-gpu-cc} compares GPU {\Read} and
{\Write} performance under the four combinations.
Two observations stand out. First, in this streaming test, GPU {\Read} is largely insensitive to both
policies, whereas GPU {\Write} varies visibly across configurations. Second,
the cost of two-way CC is paid only when the GPU L2 actually caches the data:
under WB, two-way CC becomes noticeably slower than one-way CC once the buffer
exceeds 32\,MB, matching the trend in Figure~\ref{fig:app-spark-transfer};
under UC, the two CC policies are indistinguishable.

\begin{figure}[t]
  \vspace{2mm}
  \begin{minipage}{1.\linewidth}
    \centering\includegraphics[width=0.73\linewidth]{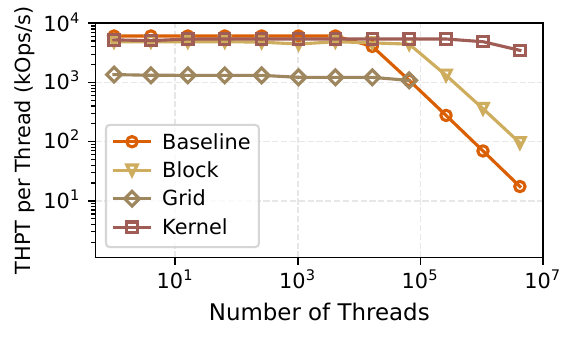}
  \end{minipage} \\[2pt]
  \begin{minipage}{1.\linewidth}
    \caption{\small{\emph{
    GPU \texttt{R${_\text{B}}$} performance as thread count increases
    under different concurrency-control approaches on Spark.
  }}}
    \label{fig:app-spark-contention}
  \end{minipage}  \\[-4pt]
\end{figure}

We hypothesize that this pattern is a direct consequence of implementing
two-way CC using the AMBA CHI protocol. In CHI terminology, one-way CC
corresponds to attaching the GPU as an IO-coherent request node (RN-I,
equivalent in role to an ACE-Lite agent): the GPU can issue snoopable requests
that invalidate CPU caches, but its own cache is not visible to the fabric and,
per the CHI specification, does \emph{not} receive snoops. Two-way CC instead
attaches the GPU as a fully coherent request node (RN-F), which by definition
holds a hardware-coherent cache, must respond to all snoop transactions, and is
registered in the Home Node's directory. 
Under WB, every cache line that enters
the GPU L2 therefore incurs directory registration on fill, ownership
acquisition before a store can retire, and de-registration on eviction---traffic
that one-way CC avoids entirely because the GPU L2 is not a participant in the
coherence domain. Under UC, the GPU L2 is bypassed, so no cache-line transitions
occur and neither CC policy pays this cost.

\stitle{GPU contention.}
Figure~\ref{fig:app-spark-contention} shows how Spark handles GPU-side
contention when many GPU threads poll the same shared location through
\texttt{R${_\text{B}}$}. The \emph{baseline} without barriers remains stable at
small and moderate thread counts, but eventually collapses as concurrent
requests concentrate on one memory location. Barrier-based throttling is
therefore still necessary: \emph{block-level barriers} reduce thread-level contention
with low fixed overhead, while \emph{grid-level barriers} remove almost all concurrent
polling at the cost of stronger synchronization requirements. On Spark,
however, the \emph{kernel-level approach} has a much smaller fixed overhead than
on Orin, so it becomes the preferred implementation for
high-contention polling. \emph{Block-level barriers} remain the next-best choice when
launching a separate polling kernel is undesirable or the polling logic must
remain inside the original kernel.

\subsection{Case Study}

We reran the Orin-based case studies on Spark through
the same platform-independent {\sys} interfaces. No Spark-specific application
code is needed once the backend exposes the required operations.

\begin{figure}[tb]
  \vspace{3mm}
  \begin{minipage}{1.\linewidth}
    \centering\includegraphics[width=0.73\linewidth]{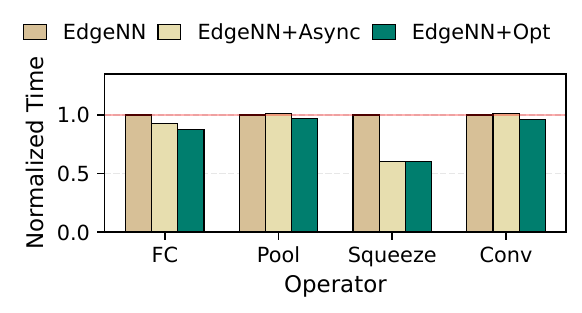}
  \end{minipage} \\[2pt]
  \begin{minipage}{1.\linewidth}
    \caption{\small{\emph{
    Control-plane coordination for hybrid inference on Spark.
  }}}
  \label{fig:app-spark-hetero}
  \end{minipage}
\end{figure}

\stitle{Hybrid inference.}
Figure~\ref{fig:app-spark-hetero} shows that the Orin optimization for hybrid
inference also applies to Spark: replacing EdgeNN's runtime-based coordination
with {\sys}'s direct-access \texttt{wait}/\texttt{set} signaling reduces
normalized operator time by 14.7\% on average relative to EdgeNN.
The gain is smaller than on Orin because Spark's lower kernel launch overhead
leaves less control-plane overhead for the signaling optimization to remove.

\begin{figure}[t]
  \vspace{2mm}
  \begin{minipage}{1.\linewidth}
    \centering\includegraphics[width=\linewidth]{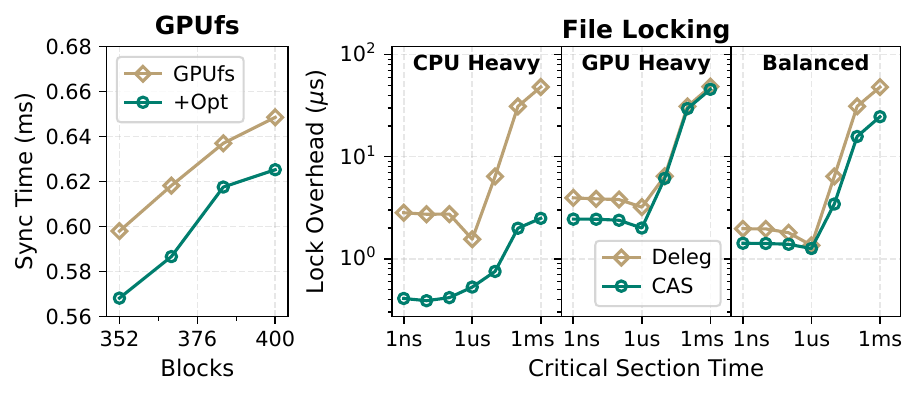}
  \end{minipage} \\[2pt]
  \begin{minipage}{1.\linewidth}
    \caption{\small{\emph{
    Control-plane coordination for GPUfs and file locking on Spark.
    Deleg denotes the delegation lock in the right-hand plot.
  }}}
  \label{fig:app-spark-fs}
  \end{minipage}  \\[-5pt]
\end{figure}

\stitle{GPUfs and file locking.}
Figure~\ref{fig:app-spark-fs} shows that the Orin control-plane optimizations
for GPU file access also carry over to Spark. In GPUfs, kernel-wide file
operations still create GPU-side polling contention, so replacing \emph{block-level}
throttling with \emph{grid-level barriers} lowers synchronization time by up to
5.1\% at high block counts. For file locking, replacing the GPU-resident
\textit{delegation lock} with a direct-access \textit{CAS lock} lets the CPU
and GPU coordinate through shared-memory atomics instead of RPC. This reduces
lock overhead by about 85\%, 37\%, and 26\% in the CPU-heavy, GPU-heavy, and
balanced configurations, respectively.

\end{document}